\documentclass[journal]{IEEEtran}

\usepackage{epstopdf}
\usepackage{amssymb,amsmath}
\usepackage{amsthm}
\usepackage{textcomp}       
\usepackage{indentfirst} 
\usepackage{url}
\usepackage{color}
\usepackage{marvosym}
\usepackage{algorithmic}
\usepackage[skip=10pt]{caption}
\usepackage{tabularx}
\usepackage{multirow}
\usepackage{hhline}
\usepackage[table]{xcolor}
\usepackage{cite}
\usepackage{setspace}
\usepackage{afterpage}
\usepackage{enumitem}
\usepackage{braket}
\usepackage{bbm}
\usepackage{bm}
\usepackage{mdframed}
\usepackage{epstopdf}
\usepackage{tcolorbox}
\usepackage{booktabs}

\usepackage{graphicx}
\usepackage{bm}
\usepackage{epstopdf}
\usepackage{subcaption}
\usepackage[ruled,commentsnumbered]{algorithm2e}

\newcounter{subeqn} %
\makeatletter
\@addtoreset{subeqn}{equation}
\makeatother

\newtheorem{theorem}{Theorem}
\newtheorem{lemma}{Lemma}

\newtheorem{assumption}{Assumption}

\usepackage{soul}

\def\1{\mathbf{1}}

\newcommand{\rev}[1]{{\color{black}{#1}}}
\newcommand{\revv}[1]{{\color{black}{#1}}}
\newcommand{\revgi}[1]{{\color{black}{#1}}}

\title{An Optimization Framework for Enabling Edge-Augmented Mobile Analytics}
\title{Selective Edge Computing for Mobile Analytics} 

\author{\IEEEauthorblockN{Apostolos Galanopoulos\IEEEauthorrefmark{1}, George Iosifidis\IEEEauthorrefmark{2}, \\ Theodoros Salonidis\IEEEauthorrefmark{3}, Douglas J. Leith\IEEEauthorrefmark{1}}\\
	\vspace{6mm}
	\IEEEauthorblockA{
		\IEEEauthorrefmark{1}School of Computer Science and Statistics, Trinity College Dublin\\
		\IEEEauthorrefmark{2}Delft University of Technology, The Netherlands\\
		\IEEEauthorrefmark{3}IBM T. J. Watson Research Center}
	\thanks{\rev{Part of this work has appeared in the Proceedings of IEEE ICC 2020 \cite{ICC_2020}.}}%
}

\IEEEoverridecommandlockouts
\begin{document}
\maketitle
\thispagestyle{plain}
\pagestyle{plain}

\vspace{-15mm}
\begin{abstract}
An increasing number of mobile applications rely on Machine Learning (ML) routines for analyzing data. Executing such tasks at the user devices saves the energy spent on transmitting and processing large data volumes at distant cloud-deployed servers. However, due to memory and computing limitations, the devices often cannot support the required resource-intensive routines and fail to accurately \revgi{execute the tasks}. In this work, we address the problem of edge-assisted analytics in resource-constrained systems by proposing and evaluating a \revgi{rigorous selective offloading} framework. The devices execute their tasks locally and outsource them to cloudlet servers only when they predict a significant performance improvement. \revgi{We consider the practical scenario where the offloading gain and resource costs are time-varying; and propose an online optimization algorithm that maximizes the service performance without requiring to know this information}. \revgi{Our approach relies on an approximate dual subgradient method combined with a primal-averaging scheme, and works under minimal assumptions about the system stochasticity. We fully implement the proposed algorithm in a wireless testbed and evaluate its performance using a state-of-the-art image recognition application, finding significant performance gains and cost savings.}

\end{abstract}


\section{Introduction}

\subsection{Background and Motivation}

The recent demand for mobile machine learning (ML) analytic applications, such as image recognition, natural language translation and health monitoring, has been unprecedented \cite{tiropanis-survey}. These services collect data streams generated by hand-held or other Internet of Things (IoT) devices, and analyze them locally or at cloud servers. The challenge with such services is that they are both resource intensive and delay sensitive. On the one hand, the cloud offers powerful ML models and abundant compute resources, but requires data transfers which consume network bandwidth and device power, as well as induce significant delays, e.g., due to intermittent connectivity \cite{cisco-cloud}. On the other hand, executing these services directly at the devices, as in \cite{microsoft-holo}, economizes network bandwidth but degrades their performance due to the devices' limited resources. For example, these nodes may have insufficient memory to support accurate deep-learning neural networks.

A promising approach to tackle this problem is to follow a middle-ground solution where the devices outsource their tasks to nearby cloudlets \cite{Cloudlets}. These edge servers are typically deployed in locations close to cellular base stations or Wi-Fi access points, and hence are in proximity with the users. Therefore, they can increase the service performance by \emph{augmenting} the devices' ML components with more accurate models, while offering tolerable communication and execution delay. Nevertheless, the success of such solutions requires intelligent decision algorithms for selecting which tasks from each device will be outsourced in order to maximize the aggregate accuracy. This is a new problem that raises intricate challenges for the network and the involved computations.

Namely, the cloudlets, unlike the cloud, have limited computing capacity and hence cannot support the requests from all devices. If overloaded, they will eventually become unresponsive. At the same time, task execution often involves the transfer of large data volumes. This calls for prudent transmission decisions in order to avoid wasting the energy of devices and congesting the network when link bandwidth is also a bottleneck. Furthermore, unlike general computation offloading solutions \cite{letaief-edge-tutorial}, in ML analytics it is imperative to identify and outsource only the tasks which can significantly benefit from cloudlet execution. Otherwise, the system will spend resources only to gain marginal performance improvements. Finally, these decisions need to be made in a dynamic fashion accounting for the time-varying network conditions, user requests and cloudlet availability; and the statistical properties of these random parameters are unknown in practice.

Our goal is to design and evaluate an online decision framework that supports edge-augmented mobile analytic services. While prior works have studied the problem of offloading computation-intensive tasks and others proposed system architectures for mobile analytics, see. Sec. \ref{sec:related}, we lack an analytical framework for maximizing the performance of such services under resource (un)availability, and time-varying network conditions. Our solution works under such practical limitations (which we measure experimentally) and is general enough to be applied to different architectures and services.

\subsection{Methodology and Contributions} 
In detail, we consider a service where a cloudlet improves upon request the execution quality of data analytic tasks that are generated by small user devices. We use as an exemplary service the processing of image frames captured by nodes such as wireless IoT cameras or small robots, that need to be processed for classifying objects of interest. Each device has a low-precision classifier, \revgi{while the cloudlet can possibly execute the task with higher precision}. The devices classify the received objects upon arrival and decide whether to transmit them to the cloudlet for further processing. This decision requires an assessment of the potential performance gains, which are measured in terms of accuracy improvement. To this end, we propose the usage of a \emph{predictor} that is installed at each device and leverages the local classification results.

In terms of resource constraints, we focus on power consumption, a bottleneck issue in small devices; and the computing capacity of the cloudlet which - unlike the cloud - is finite. The former couples the decisions of each device across time, while the latter ties the decisions of all devices sharing the cloudlet. We consider the practical case where resource availability is unknown and possibly time-varying and we observe their instantaneous values. We aim to design an algorithm that enables the coordination of devices and dictates the task outsourcing policy by carefully tuning the trade-off between maximizing the  aggregate analytics accuracy and constraining resource consumption.

We formulate the system operation as an optimization program with unknown parameters appearing both in the objective (performance gains) and constraints (power and capacity), which are learned in an online fashion. This program is decomposed via Lagrange relaxation to device-specific problems and this enables its distributed solution through an \emph{approximate} -- due to the unknown parameters -- dual ascent method. Leveraging the $\epsilon$-(sub)gradient information that is produced in the dual space by each device, we calculate primal solutions which are applied in real time. Our approach is inspired by primal averaging schemes for \emph{static} problems \cite{nedic-subgrad-siam, lindberg}, and yields a \emph{tunable} optimality bound compared to the hypothetical benchmark policy that has access to an oracle. The designed algorithm is lightweight in terms of communication overheads and adapts to resource availability and user requests. Importantly, it offers deterministic performance bounds (i.e., for each sample path) and works under minimal assumptions for the stochastic perturbations of the resources and task requests. This is in contrast with extensively-used stochastic optimization toolboxes which presume i.i.d. or Markov-modulated perturbations and offer only average guarantees, see \cite{tassiulas-book} and references therein.

Finally, the framework is extended for when the bottleneck is the wireless link capacity and for services that optimize jointly the accuracy and execution delay. \revgi{Other scenario-specific amendments are also possible, e.g., considering multi-stage services, multiple cloudlets, or other related constraints such as the cloudlet energy budget}. Given that this is a new problem, \rev{we investigate experimentally} its properties in a wireless testbed; and assess our algorithm using real datasets \cite{lecun1998gradient, Krizh2009Learn} and carefully selected benchmarks. Hence, the contributions of this work are the following: 
\begin{itemize}[leftmargin=3mm]
\item \underline{\emph{Edge-Augmented Analytics}}. We introduce the novel problem of augmenting the performance of mobile analytics using edge infrastructure (e.g., cloudlets), which is increasingly important for mobile computing services and IoT networks. \revgi{Our model can be tailored to different system architectures, types of analytic services, and resource constraints}.

\item \underline{\emph{Decision Framework}}. A task outsourcing policy is proposed that achieves near-optimal performance while being oblivious to the system's statistics. \revgi{We fully characterize the performance of the algorithm, i.e., its optimality gap, as a function of the system parameters, perturbations and the employed step rule.  To the best of our knowledge, our algorithm is the first to offer deterministic performance bounds with discrete actions, under such general conditions}; and this is a result of independent interest.

\item \underline{\emph{Implementation \& Evaluation}}. The solution is evaluated in a wireless testbed with a typical ML service and real datasets. We show that our algorithm can be  implemented as a lightweight protocol, increasing the task accuracy (up to $15\%$) and concurrently reduce the energy costs (down to $50\%$) compared to several benchmarks.
\end{itemize}
\revgi{Concluding, this work proposes a new problem, designs a novel optimization algorithm which is tailored to its needs, and uses a fully-fledged implementation in a wireless testbed in order to evaluate the proposal.}

\textbf{Organization}. Sec. \ref{sec:model} introduces the model and problem, Sec. \ref{sec:algorithms} presents the algorithm and Sec. \ref{sec:perfom} analyzes its performance. We discuss practical extensions in Sec. \ref{sec:extensions} and Sec. \ref{sec:evaluation} presents the system implementation, a series of experiments and trace-driven simulations. We discuss related work in Sec. \ref{sec:related} and conclude in Sec. \ref{sec:conclusions}. \revgi{The proofs of the various lemmas can be found in the Appendix, Sec. \ref{sec:appendix}.}

\section{Model and Problem Formulation} \label{sec:model}

We introduce the system model, the problem and the respective mathematical program. Table \ref{Tab:notation} summarizes the key notation we use throughout the paper. We use calligraphic capital letters for sets, bold typeface letters for vectors, and $\|\cdot\|$ denotes the Euclidean norm.

\subsection{Task Model}

Time is slotted and we index the slots. There is a set $\mathcal{C}$ of $C$ disjoint object classes and a set $\mathcal{N}$ of $N$ devices. Each device $n$ may receive at slot $t$ an object $s_{nt}\in \mathcal S$ for classification, where $\mathcal{S}$  is the set of possible objects, e.g., images captured by its camera. In case a device $n$ does not produce an image in slot $t$ (no task), we set $s_{nt}=\emptyset$. {Every device $n$ is equipped with a \emph{local} classifier $J_n$ which outputs the inferred class $J_n(s_{nt})\in \mathcal C$ of object $s_{nt}$ and a normalized \emph{accuracy} (or, confidence) value $d_n(s_{nt})\in[0,1]$ for that inference}\footnote{The classifier might output only the class with the highest confidence or a vector with the confidence for each class and allow the user to decide -- typically selecting the more likely class. Our analysis works for both cases.}. There is also a classifier $J_0$ at the cloudlet  which can classify any object $s_{nt}\!\in\!\mathcal{S}$ with confidence $d_0(s_{nt})$. The local classifiers may have different performance, e.g., due to possibly different ML components or training datasets, while the cloudlet classifier has the highest accuracy, i.e., $d_0(s_{nt})\!\geq\! d_n (s_{nt}), \forall s_{nt}\in S$. Parameter $\phi_{nt}\in[0,1]$ denotes the accuracy improvement when the cloudlet classifier is used:
\begin{equation}
	\phi_{nt}(s_{nt})=d_{0}(s_{nt}) - d_n(s_{nt}),\,\,\,\,\, \forall \, s_{nt}\in\mathcal{S}. \notag
\end{equation}
\revgi{It is worth stressing that several services, e.g., see YOLO \cite{yolo} , provide in real-time feedback on the confidence about the accuracy of inferences, without requiring labeled data, which exhibit indeed strong correlation with the actual accuracy \cite{automl}.} Finally,  every device is also equipped with a predictor\footnote{This can be a model-based or model-free solution, e.g., a regressor or a neural-network; our analysis and framework work for any of these solutions. In the implementation we used a mixed-effects regressor \cite{regression-multilevel}. } that is trained with the outcomes of the local and cloudlet classifiers. This predictor can estimate the cloudlet's improvement $\tilde \phi_n(s_{nt})$ for each object $s_{nt}$, where this assessment might not be exact, i.e., $\tilde  \phi_n(s_{nt})\neq \phi_{nt}(s_{nt})$; and we denote with $\sigma_{nt}(s_{nt})\in [0,1]$ the normalized predictor confidence.

\begin{table*}
\small
\begin{center}
\begin{tabular} {|l |c |}
\hline	
\textbf{Description} & \textbf{Parameter / Variable} \\ \hline
Classification confidence of $n$ (cloudlet) & $d_n\,\,\,\,(d_0)$ \\ \hline
Actual (predicted) offloading improvement & $\phi_{nt}\,\,\,\,(Q_{n})$ \\ \hline
Average power (computing) constraint & $B_n\,\,\,\,(H)$ \\ \hline
Power (Computing) resource consumption of task $s_{nt}$ at state $j$ & $o_n^j\,\,\,\,(h_n^j)$ \\ \hline
Improvement gain for device $n$ at slot $t$ (quantized value at state $j$) &  $w_{nt}\,\,\,\,(w_n^j)$ \\ \hline
Probability of system being in state $j$ & $\rho^j$ \\ \hline			
Outsourcing probability for task in state $j$ & $y_{n}^j\in[0,1]$ \\ \hline						
\end{tabular}
\end{center}
\vspace{-4mm}
\caption{Key Parameters and Variables.}
\label{Tab:notation}
\end{table*}
\normalfont

\subsection{Wireless System}

\begin{figure}[t]
	\centering
	\includegraphics[width=\columnwidth]{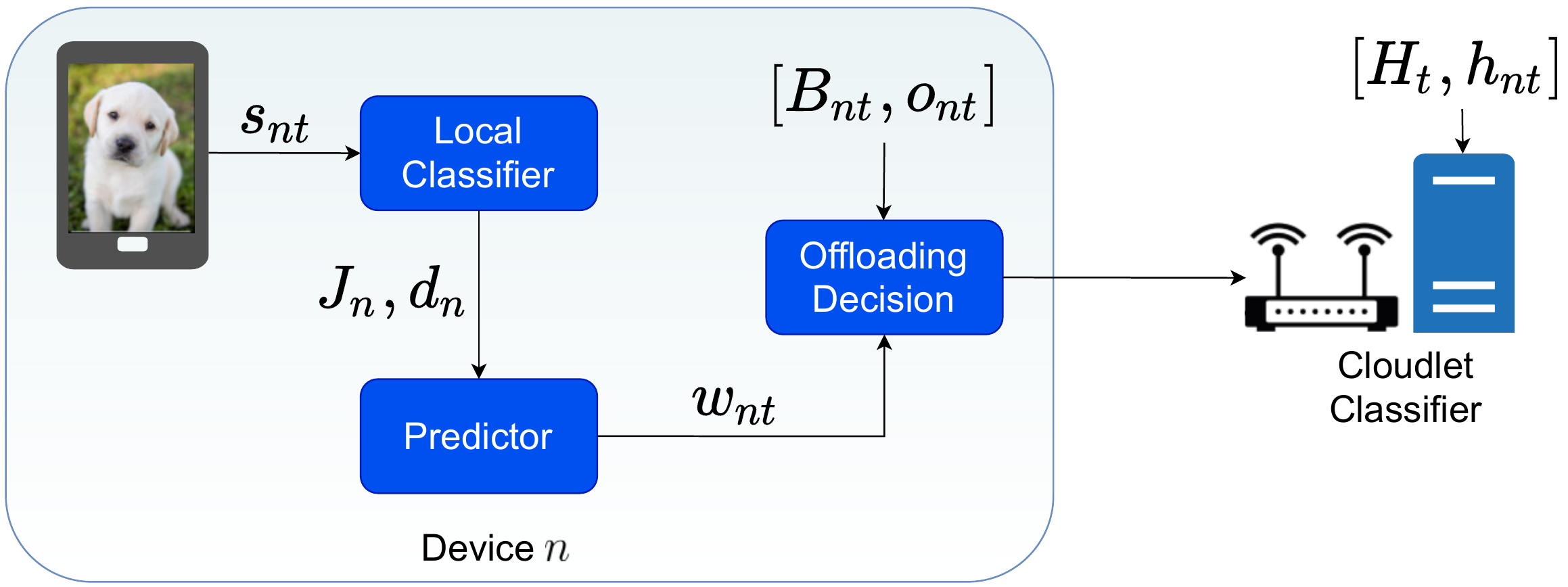}
	\vspace{-2mm}
	\caption{ \small System model including the local/cloudlet classifiers and predictors. Each device is constrained by its average power budget, and the cloudlet has a limited computation capacity.}
	\label{fig:system-model}
\end{figure}

The devices access the cloudlet through high-capacity cellular or Wi-Fi links, see Fig.~\ref{fig:system-model}, that do not impose data transfer constraints (we relax this assumption in Sec. \ref{sec:extensions} ). Each device $n$ has an \emph{average power budget} of $B_n$ Watts that it can spend on transmitting the images to cloudlet.\footnote{Local classifications can induce non-negligible energy costs to devices but these are not considered for $B_n$ since every object undergoes local classification anyway.} Average power consumption is a key limitation in such systems \cite{neely-energy}, because: the devices might have a small energy budget to spend during; their small form-factor imposes power consumption limitations; there are protocol-induced transmission constraints; or users might impose constraints on the power consumption of this service.  Similarly, the cloudlet has an \emph{average processing capacity} of $H$ cycles/sec. This resource is shared by all devices and when the total load exceeds $H$ the task delay increases fast, eventually rendering the system non-responsive.

When an image that is transmitted in slot $t$ from device $n$ to cloudlet, it consumes $o_{nt}$ Watts of the device's power budget. This quantity might change across slots due to channel conditions variations, shadowing effects, \revgi{interference from other transmission, and so on.; and follows a random process $\{o_{nt}\}_{t=1}^{\infty}$, where $o_{nt}\in \mathcal O=\{o_n^1, \ldots, O_n^{|\mathcal O|} \}$ is drawn from a set of possible values}. Also, each transmitted image $s_{nt}$ requires a number of cloudlet processing cycles $h_{nt}$, \revgi{which might vary with time, e.g., due to different image sizes, and possibly stems from a random process $\{h_{nt}\}_{t=1}^{\infty}$, with $h_{nt}\in \mathcal{H}=\{h_n^1, \ldots, h_n^{|\mathcal{H}|} \}$}. We also define $\bm{o}_t=(o_{nt}, \,n\in\mathcal{N})$ and $\bm{h}_t\!=\!(h_{nt}\leq,\, n\in\mathcal{N})$. Our model is general as the requests, power and computing costs per request can be arbitrarily time-varying and with unknown statistics.

The devices wish to involve the cloudlet only when they expect high classification precision gain with high confidence. \revgi{When the cloudlet does not offer high-enough gains or, even worse, lower accuracy, the devices need to refrain from offloading their tasks.} Otherwise, they risk consuming the cloudlet's capacity and their own power without significant benefits. Therefore, the outsourcing decision for each object $s_{nt}$ is based on the \emph{weighted improvement gain}\footnote{ \revgi{Whenever the cloudlet has lower expected accuracy from the device, then we set $w_{nt}=0$, and decide not to offload.}}: 
\begin{equation} \label{eq:predictor-gain}
	w_{nt}(s_{nt})=\tilde \phi(s_{nt})-v_n\sigma_{nt}(s_{nt}),
\end{equation} 
where $v_n\geq 0$ a \emph{risk aversion} parameter set by the system designer or each user. For example, assuming normal distribution for the outputs of the predictor we could set $v_n=1$ and use a threshold rule of $1$ standard deviation. The improvement gains follow an unknown random process $\{w_{nt}\}_{t=1}^{\infty}$, where\footnote{We note that most systems use such quantized values for the prediction gains, and the number of possible values depends on the granularity.} $w_{nt}\in \mathcal W=\{ w_n^1, \ldots, w_n^{|\mathcal W|} \}$.

\subsection{Problem Definition and Assumptions}

Our goal is to maximize the long-term accuracy improvement gains for all devices while satisfying the average capacity constraints. Let us first define the set of possible system states
\[
\mathcal J=\mathcal O^{N} \times \mathcal H^N \times \mathcal W^N,
\]
and introduce parameter $\pi_t\!\in\! \mathcal J$ that indicates the system state at slot $t$. We assume the system operation can be described by the stationary probability distribution $\bm \rho \!=\! (\rho^j, j=1, \ldots, M)$, where $M\!=\!|\mathcal J|$. We introduce variables $y_n^j \in [0,1],\ \forall n\!\in\!\mathcal N, j\!\in\! \mathcal J$ that indicate the outsourcing probability of objects from each device $n$ when the system is in state $j$. We also define the vector $\bm{y}\!=\!( y_{n}^j\!:\! n\!=\!1,\ldots,N, j\!=\!1,\ldots,M)$ and the set $\mathcal Y=[0,1]^{NM}$. Henceforth we use superscript $j$ to indicate the values of the  random variables when $\pi_t\!=\!j$.
	
Putting the above together, our optimization goal can be expressed with the following program:
\begin{align}
\mathbb P_1:\qquad&\underset{\bm y\in \mathcal Y}{\text{maximize }} \sum_{j=1}^{M}\sum_{n=1}^N w_n^jy_{n}^j\rho^j\\
&\quad\,\,\text{s.t.}\quad \,\,\,\,\, \sum_{j=1}^M y_{n}^jo_{n}^j\rho^j \le B_n,\,n\in\mathcal{N}, \label{eq:P1-device-constraint}\\
& \qquad\,\,\,\quad\,\,\,\,\, \sum_{j=1}^M\sum_{n=1}^N y_{n}^jh_{n}^j\rho^j\le H. \label{eq:P1-cloudlet-constraint}
\end{align}
Constraints \eqref{eq:P1-device-constraint} impose the average power budget\footnote{To capture the total power consumption we should add a term related to the computation energy cost at the LHS of \eqref{eq:P1-device-constraint}. However, this term is independent of the decision variable $y_{n}^j$, since the local classifier is used either way and thus it is omitted.}  of each device and \eqref{eq:P1-cloudlet-constraint} bounds the cloudlet utilization. Additional constraints can be included if needed; and we can also replace the linear objective with any other convex function. For instance, we might wish to enforce a fairness criterion by using a type of $\alpha$-fair functions \cite{fairness} or an objective that maximizes accuracy while minimizing the total delay. We elaborate on these extensions in Sec. \ref{sec:extensions}. Finally, $\mathbb P_1$ can also account for time varying capacities as we can replace $B_n$ with the time average term $\sum_j B_n^j\rho^j$, and similarly for parameter $H$; it suffices to augment the state space $\mathcal J$ accordingly.

\revgi{An important comment is in place here. If one knew in advance the value of $\bm \rho$, then we could solve $\mathbb P_1$ to obtain the optimal offloading solution $\bm y^\star$. This solution can then be implemented as a randomized policy to maximize the performance of the service}. Namely, in each slot $t$ we observe the state $\pi_t$ \revgi{and decide to offload or not based on the respective element of $\bm y^\star$}. Nevertheless, in practice one does not have access to $\bm \rho$ and hence cannot devise that optimal static policy, \revgi{i.e., cannot solve problem $\mathbb P_1$}. In line with the standard approach in stochastic optimization, cf. \cite{tassiulas-book}, we will use the unknown solution of $\mathbb P_1$ as the performance benchmark that our online algorithm aims to meet while being oblivious to the task statistics and the system parameters.

\section{Decision Framework and Online Algorithm} \label{sec:algorithms}

Our solution approach is the following: we replace the unknown parameters in $\mathbb P_1$ with their running averages, which we calculate in runtime; and we solve the modified problem with approximate gradient ascent in the dual space and perform primal averaging. \revgi{This gives us an online policy that can be implemented in real time, namely an algorithm that outputs discrete decisions, while offering performance guarantees.}

\vspace{-2mm}
\subsection{Problem Decomposition and Algorithm}

To streamline presentation we define the functions:
\[
f(\bm y)\!=-\!\sum_{n=1}^N\sum_{j=1}^M w_n^j\rho^j y_n^j ,\ g_n(\bm y)\!=\!\sum_{j=1}^M o_n^j\rho^j y_n^j - B_n, \forall n\!\in\mathcal N, 
\]
\[
g_{N+1}(\bm y)\!=\!\sum_{n=1}^N\sum_{j=1}^M h_n^j \rho^j y_n^j - H,
\]
that appear in problem $\mathbb P_1$, and we further collect all constraints in function $g(\bm y):\mathbb R^{NM}\rightarrow \mathbb R^{N+1}$. Since we can only observe the current system state, we define the respective $t$-slot functions that aggregate this information up to $t$:
\[
f_t(\bm y_t)\!=\!-\!\sum_{n=1}^N\sum_{j=1}^M y_{nt}^j w_n^j \rho_t^j ,\,\,\, g_{nt}(\bm y_t)\!=\!\sum_{j=1}^M y_{nt}^j o_n^j\rho_t^j - B_n,
\]
\[
g_{N+1,t}(\bm y_t)\!=\!\sum_{n=1}^N\sum_{j=1}^M y_{nt}^jh_n^j \rho_t^j  - H
\]
where $\rho_t^j=1/t\sum_{\tau=1}^t \mathbf{1}_{\{\pi_\tau=j\}}$ measures the distribution of state $j$ up to slot $t$ and serves as a prediction for the respective $\rho^j$ parameter. Our goal is to use the above \emph{proxy} functions in order to find a dynamic policy $\{\bm y_t\}_{t=1}^{T}$ such that the realized performance $\sum_{t=1}^Tf(\bm y_t)/T$ approaches $f(\bm y^\star)$, and similarly the induced constraint violation $\sum_{t=1}^Tg(\bm y_t)/T$ approaches $g(\bm y^\star)\preceq 0$, for any value of time horizon $T$.

The $t$-slot functions can be expressed as perturbations of the actual unknown functions:
\begin{align*}
f_t(\bm y) &=-\sum_{j=1}^{M}\sum_{n=1}^N y_{n}^jw_n^j\rho^j+\sum_{j=1}^{M}\sum_{n=1}^N y_{n}^jw_n^j(\rho^j- \rho_{t}^j)\\
&\triangleq f(\bm y) + \epsilon_t(\bm y),
\end{align*}
with $\epsilon_t(\bm y) = \sum_{j=1}^{M}\sum_{n=1}^N y_{n}^jw_n^j(\rho^j- \rho_{t}^j)$. Similarly, we write: 
\[
g_t(\bm y)=g\big(\bm y) + \delta_t(\bm y\big),\ \delta_t(\bm y)=\big(\delta_{nt}(\bm y), n=1,\ldots,N+1\big), 
\]
\[
\text{where: } \delta_{nt}(\bm y)\!=\sum_{j=1}^My_{n}^jo_n^j( \rho_{t}^j - \rho^j ), \forall n\in\mathcal N,\ 
\]
\[
\delta_{{N+1,t}}(\bm y)\!=\sum_{j=1}^M\sum_{n=1}^Ny_n^jh_{t}^j(\rho_{t}^j - \rho^j )
\]
Next, we can define a new problem for each slot $t$:
\begin{align}
	\mathbb P_2(t):\quad&\underset{\bm y\in \mathcal Y}{\text{maximize }} f_t(\bm y) \quad \emph{s.t.} \quad g_t(\bm y)\preceq 0.
\end{align}
We will use $\{\mathbb P_2(t)\}_t$ to perform a dual ascent  and obtain the $\{\bm y_t\}_t$ that applied in real time. 

First, we dualize $\mathbb P_2(t)$  and introduce the Lagrangian:
\begin{equation}
	L_t(\bm y, \bm \lambda)= f_t(\bm y)+\bm \lambda g_t(\bm y) \notag
\end{equation}
where $\bm \lambda\!=\!(\lambda_1, \lambda_2, \ldots, \lambda_N, \mu)$ are the non-negative dual variables for the $N+1$ constraints. The dual function is:
\begin{equation}
	V_t(\bm \lambda)=\min_{\bm y\in\mathcal Y} L_t(\bm y, \bm \lambda).		\notag
\end{equation}
The basis of our approach is the application of a dual-ascent algorithm where the iterations are in sync with the system's time slots $t$. Specifically, in each iteration $t$ we can minimize the Lagrangian by executing:
\begin{align}
	y_n^{j, \star}=\arg\min_{y_n^j\in [0,1]} y_{n}^j \big( -w_n^j + \lambda_{nt}o_n^j + \mu_t h_n^j\big)\rho_t^j.
\end{align}
This yields the currently optimal offloading policy for each state $j\in\mathcal J$, based on which we derive an easy-to-implement offloading rule. Namely, denoting with $j_t$ the state at slot $t$, we  write for the offloading decision of each device $n\in\mathcal N$:
\begin{equation}
	y_{n}^{j_t}=
	\begin{cases} 
		1 & \text{if }\quad \lambda_{nt}o_{n}^{j_t}+\mu_t h_{n}^{j_t}< w_{n}^{j_t} \\
		0       & \text{otherwise.}
	\end{cases} \label{eq:sub1}
\end{equation}
Note that in practice, state $j_t$ is not entirely known to each device $n \in \mathcal{N}$, but it rather refers to the partial system state regarding the device. This is possible since each device knows its own expected power consumption of the current slot by, e.g. estimating the channel state and also the expected cloudlet resource consumption through the image's file size.  
Eq. \eqref{eq:sub1} dictates an offloading when the expected accuracy gain $w_n^{j_t}$ exceeds the weighted resource cost $\lambda_{nt}o_n^{j_t}+\mu_th_n^{j_t}$. 
 
Then, we improve the current value of $V_t(\bm \lambda)$ by updating the dual variables:
\begin{align}
\lambda_{n,t+1}&=\Big[ \lambda_{nt}+a_t\big( \sum_{j=1}^M o_n^j \rho_t^j y_{n}^{j} - B_n	\big) \Big]^+,\ \forall n\in\mathcal N, \label{dual-update1} \\
\mu_{t+1}&=\Big[ \mu_{t}+a_t\big( \sum_{n=1}^N\sum_{j=1}^M h_n^j \rho_t^j y_{n}^{j} - H	\big) \Big]^+  \label{dual-update2}
\end{align}
where $[u]^+=\max\{0,u\}$ and $a_t$ is the dual step.

The online task outsourcing algorithm, henceforth called \emph{OnAlgo}, is based on eq.  \eqref{eq:sub1}-\eqref{dual-update2}. The details are presented in Algorithm 1. When each device $n$ receives an object $s_{nt}$ in slot $t$, it uses its classifier to predict its class and the predictor to estimate the cloudlet's classification improvement (Steps 5-7). Then, the device uses its threshold decision rule (Step 9) that compares the expected benefits for state $j_t$ with the outsourcing costs for the device and cloudlet. \revgi{If the cloudlet is not expected to offer satisfactory gains (or, even worse, has lower accuracy), the devices refrain from offloading their tasks}. The devices receive the updated state distribution from the cloudlet (Step 12), and update their local dual variable for the power constraint (Step 13). The clouldet initially evaluates the current system state and sends it to the devices (step 15). Then, it classifies the received objects and updates its congestion variable (Step 17), which is sent to devices. 

\revgi{It is interesting to observe that OnAlgo is lightweight in its computation and communication requirements. Namely, the offloading decision are made simply by using an intuitive threshold rule that weights the expected performance gains with the expected costs  where the latter are captured in a systematic way via the dual multipliers (known also as \emph{shadow prices}). And this rule can be employed by each device independently. Similarly, the updates of the dual variables are very simple as they involve summation of scalars and projection onto the non-negative orthant, i.e., keeping only the positive result or setting equal to zero otherwise.}

\begin{algorithm}[t]
\small{	
	\caption{OnAlgo}
	\label{OnAlgo}
	\begin{algorithmic}[1]
		\STATE \textbf{Initialization:} $t=0, \bm \lambda_0\!=\!0,\ \forall\ n, j$
		\WHILE {True}
		\FOR {\underline{\textbf{each} device $n\in\mathcal{N}$}}
		\STATE $ y_{nt}^j=0,\,\forall j$
		\STATE Receive object $s_{nt}$
		\STATE \revv{Classify objects and obtain $J_n(s_{nt}),d_n(s_{nt}),\,\,\forall s_{nt}$}
		\STATE \revv{Use classification results on predictor to obtain $w_{nt}$}
		\STATE Observe partial current state $j_t$ and send it to cloudlet
		\IF {$\lambda_{nt} o_n^{j_t} + \mu_t h_n^{j_t} < w_{n}^{j_t}$}
		\STATE ${y}_{n}^{j_t} \leftarrow 1$ \,\,\,\,\% Send object to cloudlet
		\ENDIF
		\STATE Receive updated distribution $\rho_t^j$ from the cloudlet
		\STATE $\lambda_{n,t+1} \leftarrow [\lambda_{nt}+\alpha_t(\sum_{j=1}^M o_n^j \rho_t^j y_{n}^{j} - B_n)]^+,\,\forall n\in\mathcal{N}$
		\ENDFOR	
		\STATE \underline{Cloudlet:} Receive partial system states from devices, and send back $\rho_t^j$
		\STATE Compute tasks received from all devices
		\STATE $\mu_{t+1} \leftarrow [\mu_{t}+\alpha_t(\sum_{n=1}^N\sum_{j=1}^M h_n^j \rho_t^j y_{n}^{j}- H)]^+$
		\STATE Send $\mu_{t+1}$ to devices
		\STATE $t \leftarrow t+1$
		\ENDWHILE
	\end{algorithmic}}
\end{algorithm}

\section{Performance Analysis}\label{sec:perfom}
The gist of our approach is that, as time evolves, the sequence of  problems $\{ \mathbb P_2(t)\}_t$ that aggregate the statistical information up to slot $t$, approaches the original problem $\mathbb P_1$. We note that the following analysis is general as it holds for different functions $f(\bm y)$ and $g(\bm y)$ than the above, as long as they are convex. We first introduce formally the necessary assumptions and then present a set of technical Lemmas that lead to our main Theorem.

\begin{assumption}\label{asum1}
The constraint functions and the objective functions of $\{\mathbb P_2(t)\}_t$ satisfy:
$|f_t(\bm y)|\leq \sigma_f$,	$\| g_t(\bm y)\| \leq \sigma_g,\,\,\, \forall t,\,\bm y\in \mathcal Y$.
\end{assumption}

\begin{assumption}[Slater Condition]\label{asum2}
There exists a vector $\bm y_s\in \mathcal Y$ such that $g_t(\bm y_s) \prec 0, \forall t$. 
\end{assumption}

\vspace{-1mm}
\subsection{Complementary Slackness and Constraint Bounds} 
\begin{lemma}[Complementary Slackness Lower Bound]\label{lem:complower}
Under the dual update \eqref{dual-update1}-\eqref{dual-update2} it holds:	
	\begin{align}	
		-\sum_{t=1}^T \bm \lambda_{t}^\top g_t(\bm y_t) &\leq \frac{\sigma_g^2}{2}\sum_{t=1}^T a_t +\frac{1}{2}\sum_{t=1}^T\|\bm \lambda_t\|^2 \left( \frac{1}{a_t}-\frac{1}{a_{t-1}} \right) \notag\\
		&-\frac{\|\bm \lambda_{T+1}\|^2}{2a_T} \label{lemma2}
	\end{align}
\end{lemma}

The next result bounds the constraint violation of OnAlgo.
\begin{lemma}[Bounded Constraint Violation]\label{lem:viol}
Under the dual update \eqref{dual-update1}-\eqref{dual-update2} it holds:	
\begin{align}
\left\| \frac{1}{T}\sum_{t=1}^Tg(\bm y_t) \right\| &\leq \frac{\Vert \bm \lambda_{T+1} \Vert}{Ta_T} + \frac{1}{T}\sum_{t=1}^T\| \bm \lambda_{t}\|\left( \frac{1}{a_{t-1}} - \frac{1}{a_t} \right) \\ \notag
&+ \frac{1}{T}\sum_{t=1}^T \|\delta_t(\bm y_t)\|.
\end{align}
\end{lemma}

\vspace{-1mm}
\subsection{Approximate Primal Averaging Bounds} The basic idea is that OnAlgo converges to an approximate saddle point.  Approximate complementary slackness then allows us to bound the performance gap. We use the next lemma.

\begin{lemma}[Approximate Saddle Point]\label{lem:saddle}
When $\{\bm y_t\}_t$ are selected using \eqref{eq:sub1}, the $t$-slot Lagrangian is bounded by:
\begin{align}
\frac{1}{T}\sum_{t=1}^TL_t(\bm y_t, \bm \lambda_t)  - f(\bm y^\star) \le   \frac{1}{T} \sum_{t=1}^{T} \Big( \epsilon_{t}(\bm z_t)+\bm \lambda_{t}^\top \delta_t(\bm z_t)\Big)\label{eq:primal_bound}
\end{align}
where  $\bm z_t\in\arg\min_{\bm y\in \mathcal Y} f(\bm y)+\bm \lambda_t^\top g(\bm y)$.  
\end{lemma}

We can now state and prove the main theorem.
\begin{mdframed}[innerbottommargin=12pt]
\vspace{-2mm}
\begin{theorem}[Performance Bounds]\label{lemma:subgrad}
OnAlgo ensures:
\begin{align*}
&\textbf{(a)}:\,\,\,\frac{1}{T}\!\sum_{t=1}^Tf(\bm y_{t})\!-\!f(\bm y^\star) \leq C_T+\frac{\sigma_g^2}{2T}\sum_{t=1}^T a_t \\
&\quad \quad +\frac{1}{2T}\sum_{t=1}^T \|\bm \lambda_{t}\|^2\left( \frac{1}{a_t} - \frac{1}{a_{t-1}} \right) - \frac{\|\bm \lambda_{T+1}\|^2}{2Ta_T} \\
&\textbf{(b)}:\,\,\,\Big\|\frac{1}{T}\sum_{t=1}^Tg(\bm y_t) \Big\| \leq \frac{\Vert \bm \lambda_{T+1} \Vert}{Ta_T} \\
&\quad \quad+ \frac{1}{T}\sum_{t=1}^T\| \bm \lambda_{t}\|\left( \frac{1}{a_{t-1}} - \frac{1}{a_t} \right) + \frac{1}{T}\sum_{t=1}^T \|\delta_t(\bm y_t)\| \notag \\
& \text{where} \quad C_T= \frac{1}{T} \sum_{t=1}^{T} \Big( \epsilon_{t}(\bm z_t)-\epsilon_{t}(\bm y_t)+\bm \lambda_{t}^\top \delta_t(\bm z_t)\Big),\\
&\bm z_t\in\arg\min_{\bm y\in \mathcal Y} f(\bm y)+\bm \lambda_t^\top g(\bm y).
\end{align*}
\end{theorem}
\vspace{-4mm}
\end{mdframed}
\vspace{-1mm}
\begin{IEEEproof}
Replacing the definition of the $t$-slot Lagrangian, $L_t(\bm y_t, \bm \lambda_t)= f_t(\bm y_t)+\bm \lambda_t^\top g_t(\bm y_t)$, in Lemma \ref{lem:saddle} and subtracting $(1/T)\sum_{t=1}^T \bm \lambda_t^\top g_t(\bm y_t)$ from both sides we can write:
\begin{align*}
\frac{1}{T}\sum_{t=1}^T f_t(\bm y_t) - f(\bm y^\star) &\leq \frac{1}{T}\sum_{t=1}^T\left( \epsilon_t(\bm z_t)+\bm \lambda_t^\top \delta_t(\bm z_t)	\right)\\
&-\frac{1}{T}\sum_{t=1}^T \bm \lambda_t^\top g_t(\bm y_t), \notag
\end{align*}
and finally expanding $f_t(\bm y_t)=f(\bm y_t)+\epsilon_t(\bm y_t)$ and using Lemma \ref{lem:complower}, we eventually get:
\begin{align}
&\frac{1}{T}\sum_{t=1}^T f(\bm y_t) - f(\bm y^\star) \le \frac{1}{T}\sum_{t=1}^T\left(\epsilon_t (\bm z_t)-\epsilon_t(\bm y_t)+\bm \lambda_t^\top \delta_t(\bm z_t)\right)\notag \\ &+\frac{\sigma_g^2}{2T}\sum_{t=1}^Ta_t +\frac{1}{2T}\sum_{t=1}^T\|\bm \lambda_t\|^2 \left( \frac{1}{a_t}-\frac{1}{a_{t-1}} \right) -\frac{\|\bm \lambda_{T+1}\|^2}{2Ta_T}.\notag
\end{align}
The second claim of the Theorem follows from Lemma \ref{lem:viol}. 
\end{IEEEproof}

Theorem 1 characterizes the optimality gap and constraint violation for any value of the time horizon $T$. The steps can be selected either to be constant, e.g., $a_t=a$ as in \cite{nedic-subgrad-siam} or to be diminishing, e.g., $a_t=a/t^\beta$, with $\beta\in (0,1)$. Also, the theorem reveals how the error terms of the proxy functions affect the convergence; and it is valid even if one uses other types of estimators, e.g., employing Gaussian Processes to approximate the objective and constraints. 

%
%
%
%
%
%
%
%
%
%
%

\subsection{Convergence Analysis}

\revgi{The final step of our analysis is to study the convergence of the proposed algorithm. First, it is important to see  that Theorem 1 provides a full characterization of the performance gap, and demonstrates how this depends on the time horizon, the system parameters (e.g., $\sigma_g$), the system perturbations (errors), and the update steps $\{a_t\}_t$. The convergence rate depends on all these factors. We start by proving that $\bm \lambda_t$ is bounded $\forall t$, which is a technical requirement for our analysis.}

\subsubsection{Boundedness of Multipliers} For $\bm \lambda_t$ to remain bounded we need $g_t(\bm y_t)$ to converge to $0$ sufficiently quickly or to be negative sufficiently often. We start with the following result:
\begin{lemma}[Bounded level set] \label{lemma:bounded_set}
Under Assumptions \ref{asum1}-\ref{asum2} and defining $q:=\! \min_t q_t$ with $q_t\!=\! \min_n \{-g_{nt}(\bm y_s)\}\! >\!0$, it holds: 
\[
\sum_{n=1}^{N+1} \lambda_n \leq \big(\sigma_f - v\big)/q,\quad \forall\, \bm \lambda \in \mathcal Q_{v}:= \{\bm \lambda \succeq 0\ |\ V_t(\bm \lambda) \geq v\}
\]
\end{lemma}

\begin{lemma}[Dual vector bound] \label{lemma:dual_bound}
Under Assumptions \ref{asum1}-\ref{asum2}, the dual update \eqref{dual-update1}-\eqref{dual-update2} ensures $\|\bm \lambda_t\|$ is uniformly bounded.
\end{lemma}

Obtaining an upper bound for the norm $\| \bm \lambda_t\|$ ensures that the respective terms $\| \bm \lambda_T\|/T$ appearing on the bounds of Theorem 1, are guaranteed to diminish with time.

\subsubsection{Error Terms} Finally, we characterize the aggregate error terms that are induced by the employed approximate dual method, and which affect the bounds of Theorem 1. We write:
\begin{align}
\frac{1}{T}\sum_{t=1}^T\!\! \epsilon_t(\bm z_t)\!-\!\epsilon_t(\bm y_t)=\frac{1}{T}\sum_{t=1}^T \sum_{j=1}^M\sum_{n=1}^N w_n^j (z_{nt}^j\!-\!y_{nt}^j)(\rho_t^j\!-\!\rho^j) \label{eq-error}
\end{align}
and similarly the other error-related terms of the performance and constraint bounds:

\begin{align*}
\frac{1}{T}\!\sum_{t=1}^T\bm \lambda_t^\top\delta_t(\bm z_t)\!&=\!\frac{1}{T}\sum_{t=1}^T \sum_{j=1}^M\sum_{n=1}^N\! o_n^j \lambda_{nt} z_{nt}^j(\rho_t^j \!- \rho^j)\\
&+ \frac{1}{T}\sum_{t=1}^T\sum_{j=1}^M h_n^j\mu_tz_{nt}^j(\rho_t^j - \rho^j),
\end{align*}
\begin{align*}
	\frac{1}{T}\!\sum_{t=1}^T\delta_t(\bm y_t)\!&=\!\frac{1}{T}\sum_{t=1}^T \sum_{j=1}^M\sum_{n=1}^N\! o_n^j y_{nt}^j(\rho_t^j \!- \rho^j)\\
	&+ \frac{1}{T}\sum_{t=1}^T\sum_{j=1}^M h_n^j y_{nt}^j(\rho_t^j - \rho^j).
\end{align*}

Now, we can upper bound the above terms by their  norms and observe that, since parameters $w_n^j, \forall n,j$ and the offloading variables are uniformly bounded, their overall behavior depends on terms $|\rho_t^j - \rho^j|, \forall j\in\mathcal J$. Hence, as long as the running average of the realizations for each state $j$ converge to the respective mean value, the errors gradually diminish to zero. The conditions that ensure this convergence range from the random variables $\{\bm 1_{\{\pi_t=j\}}\}_{t,j}$ being i.i.d. where the Law of Large Numbers applies; to more general settings where they are independent and we can use Hoeffding's inequality \cite[Theorem 1]{hoeff} to obtain:
\begin{align}
Prob\Big(  |\rho_t^j - \rho^j |>\kappa	\Big)< \frac{1}{e^{2\kappa^2t }}.
\end{align}
And one can further relax the assumptions regarding the system state statistics, e.g., to allow for a martingale-type weakly dependence across successive states, and employ the Azuma inequality for a similar bound. These conditions generalize the stricter requirements of i.i.d. statistics that other network optimization frameworks require \cite{tassiulas-book}.

\revgi{Concluding, it is interesting to consider some special cases in order to shed light on the favorable convergence properties of our algorithm. Namely, for the case where we use the step $a_t=a/\sqrt t$, Theorem 1(a) shows that the average gap closes at a rate of $\mathcal O(1/\sqrt T)$. To see this, first note that it holds:
\begin{align}
\frac{\sigma_g^2}{2T}\sum_{t=1}^T a_t \leq  \frac{\sigma_g^2}{2T}\sum_{t=1}^T \frac{a}{\sqrt t}\leq \frac{\sigma_g^2}{2T} 2a\sqrt T=\frac{2a\sigma_g^2}{\sqrt T} =\mathcal O\big({T^{-1/2}}\big) \nonumber
\end{align}
We can also bound the next term:
\begin{align*}
&\frac{1}{2T}\sum_{t=1}^T \|\lambda_t\|^2 \left( \frac{1}{a_t}\!-\!\frac{1}{a_{t-1}} \right)\!=\!	\frac{1}{2aT}\sum_{t=1}^T \|\lambda_t\|^2 \left(\sqrt t \!- \sqrt{t-1} \right)\\ 
&\stackrel{(a)}\leq \frac{\|\lambda_{max}\|^2}{2aT}\sum_{t=1}^T \left(\sqrt t \!- \sqrt{t-1} \right)=\frac{\|\lambda_{max}\|^2}{2a\sqrt T}=\mathcal O\big( T^{-1/2}\big)
\end{align*}
where $(a)$ follows from Lemma 5 (dual vectors uniformly bounded). And similarly we can bound the last RHS term in Theorem 1(a) by $\mathcal O(T^{-1/2})$. Now it remains to bound $C_T$. Indeed, when the perturbations are i.i.d. the gap between the running average of the state probabilities $(\rho_t^j)$ and their mean values $(\rho^j)$ diminishes at the rate of $\mathcal O(T^{-1/2})$. Hence, using the fact that $\sum_{t=1}^T 1/\sqrt t\leq 2\sqrt T$, and that all variables are bounded in $[0,1]$ the error term in \eqref{eq-error} diminishes with rate $\mathcal O(1/\sqrt T)$ as well. Finally, it is easy to see that with a similar argument we find that the constraint violation diminishes with rate $\mathcal O(1/\sqrt T)$ in this case. Hence, overall the algorithm converges with that rate, both w.r.t. the optimality gap and the constraint violation.}

\section{Model and Algorithm Extensions} \label{sec:extensions}

We extend our framework by jointly optimizing prediction accuracy {and} total execution delay, since the latter can also be crucial for many edge services. \revgi{Then, we explain how it can cope with massive demand scenarios where the wireless bandwidth becomes a bottleneck or the cloudlet's energy cost is significant}; and finally we elaborate on alternative designs/usages of the predictor. 
 
\textbf{Joint Accuracy and Delay Optimization}: We extend our model to capture both the accuracy gains and the impact of offloading decisions on delay. We do so by adding the total delay for processing the tasks of all users in the objective function and using a scaling parameter $\zeta \in [0,1]$ to balance between the two objectives. In detail, we can express the total delay as:
\begin{align*}
D_{tot}(\bm y)= \sum_{n=1}^N\sum_{j=1}^M \big(1-y_n^j\big)D_n^{pr} + y_{n}^j\big( D_n^{pr} + D_0^{pr} + D_n^{tr} \big),
\end{align*}
where $D_n^{pr}$, $D_0^{pr}$ are the delays for processing images at device $n$ or the cloudlet, respectively; and $D_n^{tr}$ the delay for transmitting images to cloudlet. These quantities can vary with time, similarly to the other system parameters, because each image has different size or the wireless medium changes. 
The processing delays can be modeled with linear functions as we enforce the processing capacity constraints. Namely, we can write $D_n^{pr}=k_{n}/H_{d,n}$, where $k_n$ is the number of CPU cycles required for processing the images of device $n$, and $H_{d,n}$ is the processing speed of device $n$ (cycles/sec). Similarly, we can define the processing delay at the cloudlet as $D_0^{pr}=k_{n}/H$ which may vary with time; we refer the reader also to \cite{xu-1} and references therein. 

Regarding the transmission delay, this depends on the actual system architecture. For example, if different channels are employed for the users, we can express it as $D_n^{tr}=\ell_n/(r_nW)$, where $\ell_n$ is the size of each image, $r_n$ the channel gain for user $n$, and $W$ the link bandwidth. If there is a CSMA-type network where users need to share their links, we need to replace $W$ with the actual airtime $W_n$ that user $n$ receives; and in the case we have a fair round-robin (vanilla version of CSMA) we can approximate this with $W_n=\sum_{j=1}^My_n^j /\sum_{n=1}^N\sum_{j=1}^My_n^j$. This model has been used extensively in Wi-Fi service allocation, see \cite{{yang-wifi}}, and in mobile code offloading, e.g., in \cite{xu-1}.

Following the analysis in Sec. \ref{sec:model} we can replace in $\mathbb P_1$ the new objective function $f(\bm y)=\sum_{j=1}^{M}\sum_{n=1}^N w_n^jy_{n}^j\rho^j - \zeta D_{tot}(\bm y)$, and by following the same process obtain the offloading rule:
\begin{equation}
y_{nt}^{j_t}=
\begin{cases} 
1 &\,\,\,\,\text{if}\,\,\,\,\, \lambda_{nt}o_{n}^{j_t}+\mu_t h_{n}^{j_t}< w_{n}^{j_t} - \zeta(D_{nt}^{tr} + D_{0t}^{pr})\\
0       & \text{otherwise}
\end{cases} \label{eq:sub2},
\end{equation}
where we observe that the device execution delay is nullified since it is independent of the offloading decision, and the condition in line 10 of Algorithm \ref{OnAlgo} will be replaced by \eqref{eq:sub2}.

\textbf{Wireless Bandwidth and Energy Cost Constraints}: We have assumed the system operation is constrained by the devices' power budget and the computing capacity of the cloudlet. Indeed, most often these are the bottleneck resources \cite{Cloudlets,letaief-edge-tutorial,Smart_cities}. However, in scenarios of massive demand the wireless link capacity might also be a bottleneck constraint. Our analysis can be readily extended for this case. If we denote with $\{W_t\}_{t=1}^{\infty}$ the link capacity process (uniformly bounded; well-defined mean value \revv{$W$}) assuming a wireless link shared by all devices\footnote{This can be either an OFDM-based cellular link or a coordinated access WiFi link; in the case we have a CSMA-type of mechanism, one needs to account for the additional bandwidth loss due to collisions, etc.}, we can add to $\mathbb P_1$ the constraint:
 
\revv{
\begin{equation}
\sum_{n=1}^N\sum_{j=1}^M y_{n}^j\rho^j\ell_{n} \leq	W , \label{eq:link-constraint-new}
\end{equation}
}
where $\ell_{n}$ is the size of objects device $n$ transmits. Eq. \eqref{eq:link-constraint-new} can be handled as the computing constraint \eqref{eq:P1-cloudlet-constraint} and will only affect the convergence bounds. \revgi{Similarly, we can include other constraints that couple the actions of all devices, such as the energy cost at the cloudlet which increases with the aggregated offloaded tasks from all devices and might depend on time-varying energy prices.}

\textbf{Alternative System Architectures}: A different mechanism is possible, where the devices send objects to the cloudlet before using their own classifier. This approach can reduce the consumed energy, since it avoids low-accuracy local classifications. However, it requires a different type of a predictor, namely one that can estimate the expected accuracy gain using some basic features of the object (e.g., its file size), and without requiring input from the local classifier. In this case, modeling the power consumption of the devices would modify constraint \eqref{eq:P1-device-constraint} of $\mathbb P_1$ as:
\begin{equation}
\sum_{j=1}^M \Big( y_{n}^j\rho^jo_{n} + (1-y_{n}^j)\rho^j\nu_{n} \Big) \le B_{n},\,\,\,\forall n\in\mathcal{N}, \notag	
\end{equation}
where the second term \revv{indicates the power $\nu_n$ consumed by each device when only local classification is performed}. OnAlgo can be extended to this case by changing the predictor. 

Similarly, it is possible to have services that are executed in multiple stages, e.g., a video stream is compressed, then frames of interest are selected, and objects are identified on each frame. In this case, the devices might decide to outsource some of the tasks in the first stage, some others after the second stage, and so on. Again, our optimization algorithms can be extended to include these decisions, by defining a separate set of variables for each stage while accounting for the costs and properties (e.g., data volumes) in each case. In specific, \eqref{eq:P1-device-constraint} would be transformed to:
\begin{equation}
\sum_{j=1}^M \Big( y_{n}^j\rho^jo_{n} + (1-y_{n}^j)\rho^j\nu_{n}^{cl} \Big) \le B_{n},\,\,\,\forall n\in\mathcal{N}, \notag	
\end{equation}
where $\nu_{n}^{cl}$ is the classification computing cost, which is significantly smaller than $\nu_{n}$. Observe that the computing load of stage 1, i.e. feature extraction, is not accounted for since it is again induced regardless of the offloading decision. 
\section{Implementation and Evaluation} \label{sec:evaluation}

We have fully implemented the proposed architecture, evaluated OnAlgo with real datasets, and complemented our analysis with large-scale synthetic simulations. This section has four goals: \emph{(i)} investigate the accuracy performance of well-known classifiers for different sizes of training datasets, hence revealing why \emph{edge augmentation} is needed;  \emph{(ii)} Measure the energy and computing costs of image classification tasks; \emph{(iii)} Perform a parameter-sensitivity analysis of OnALgo; and \emph{(iv)} Compare OnALgo with several benchmark algorithms.

\subsection{Experiments Setup}
 
\subsubsection{Testbed and Measurements} 
\begin{figure*}[t!]
	\begin{subfigure}[b]{0.24\linewidth}
		\centering
		\includegraphics[width=3.9cm]{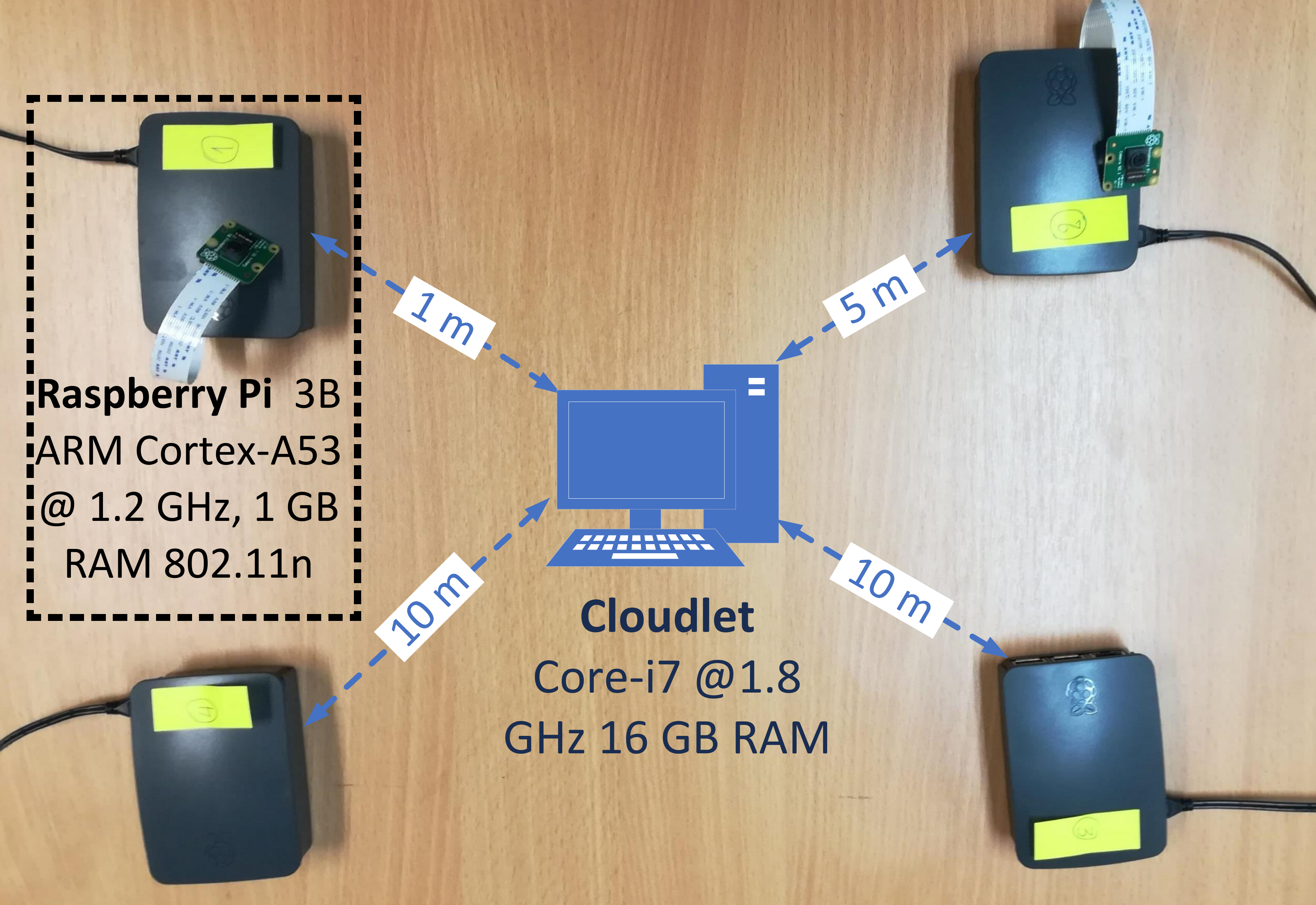}
		\caption{Testbed Layout}
	\end{subfigure}
	\begin{subfigure}[b]{0.24\linewidth}
		\centering
		\includegraphics[width=3.9cm]{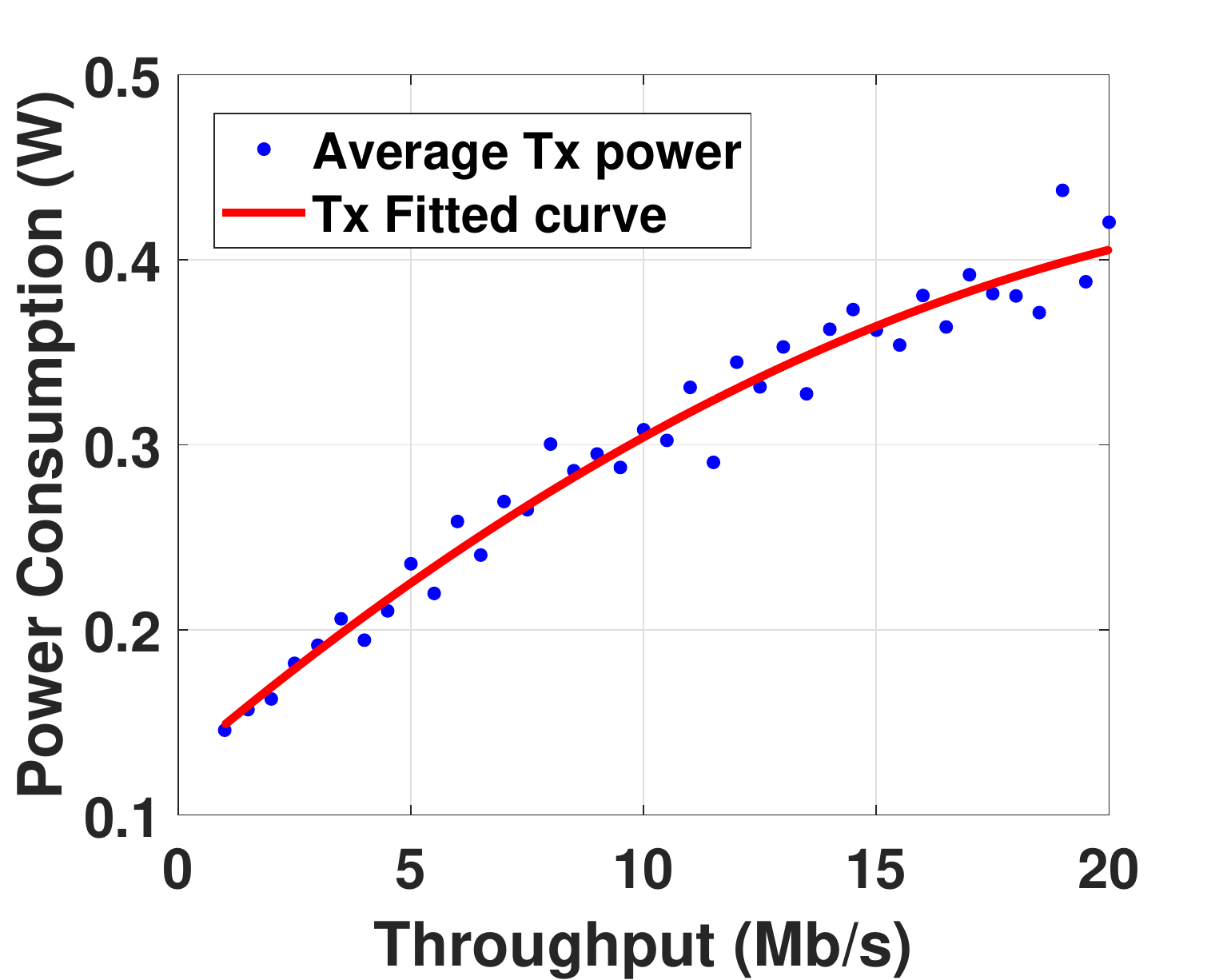}
		\caption{\small Tx power consumption}	\label{fig:Tx_power2}
	\end{subfigure}
	\begin{subfigure}[b]{0.24\linewidth}
		\centering
		\includegraphics[width=4.1cm]{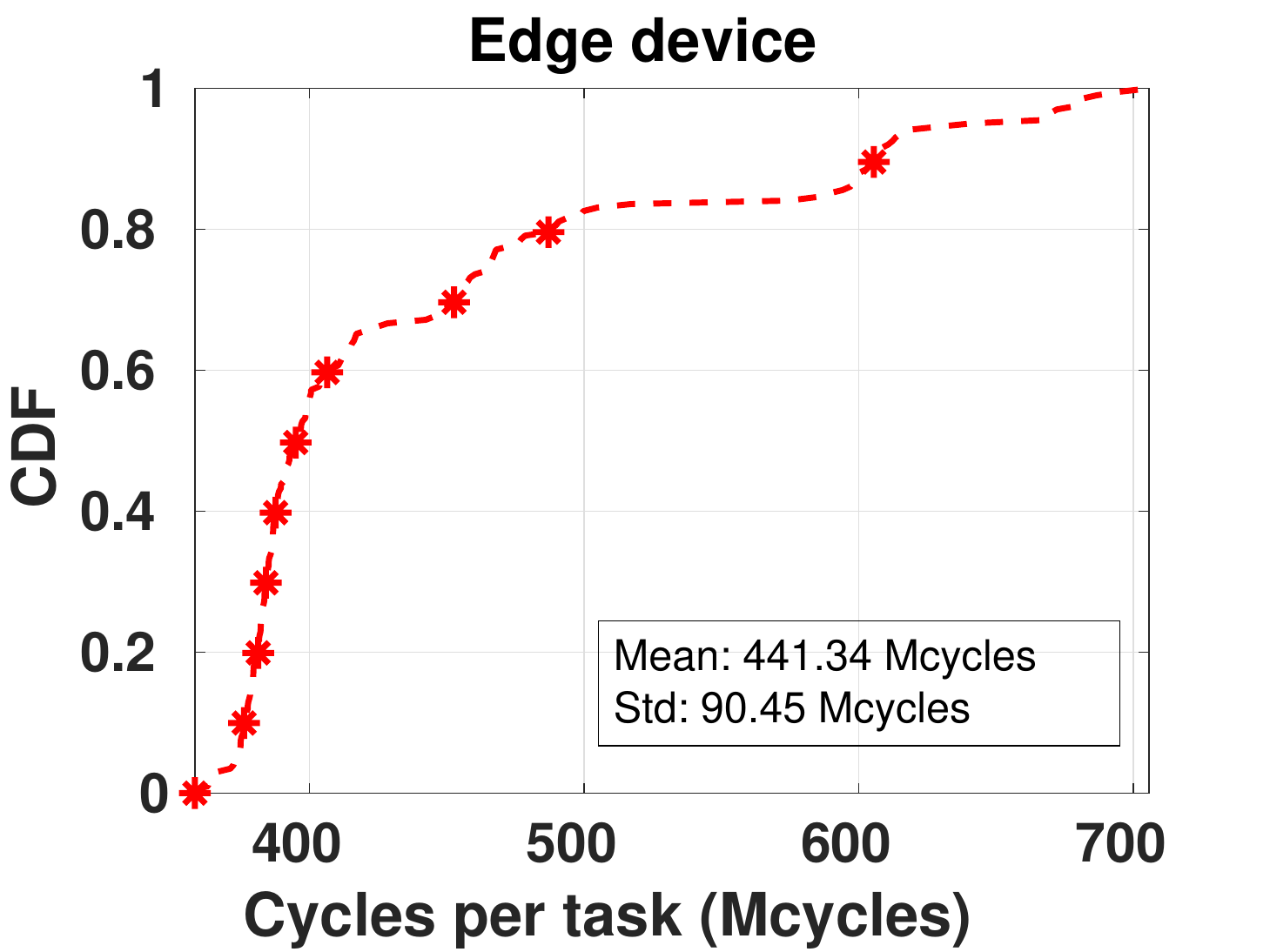}
		\caption{Cloudlet cycles per task}
		\label{fig:cycles}
	\end{subfigure}
	\begin{subfigure}[b]{0.24\linewidth}
		\centering
		\includegraphics[width=4cm]{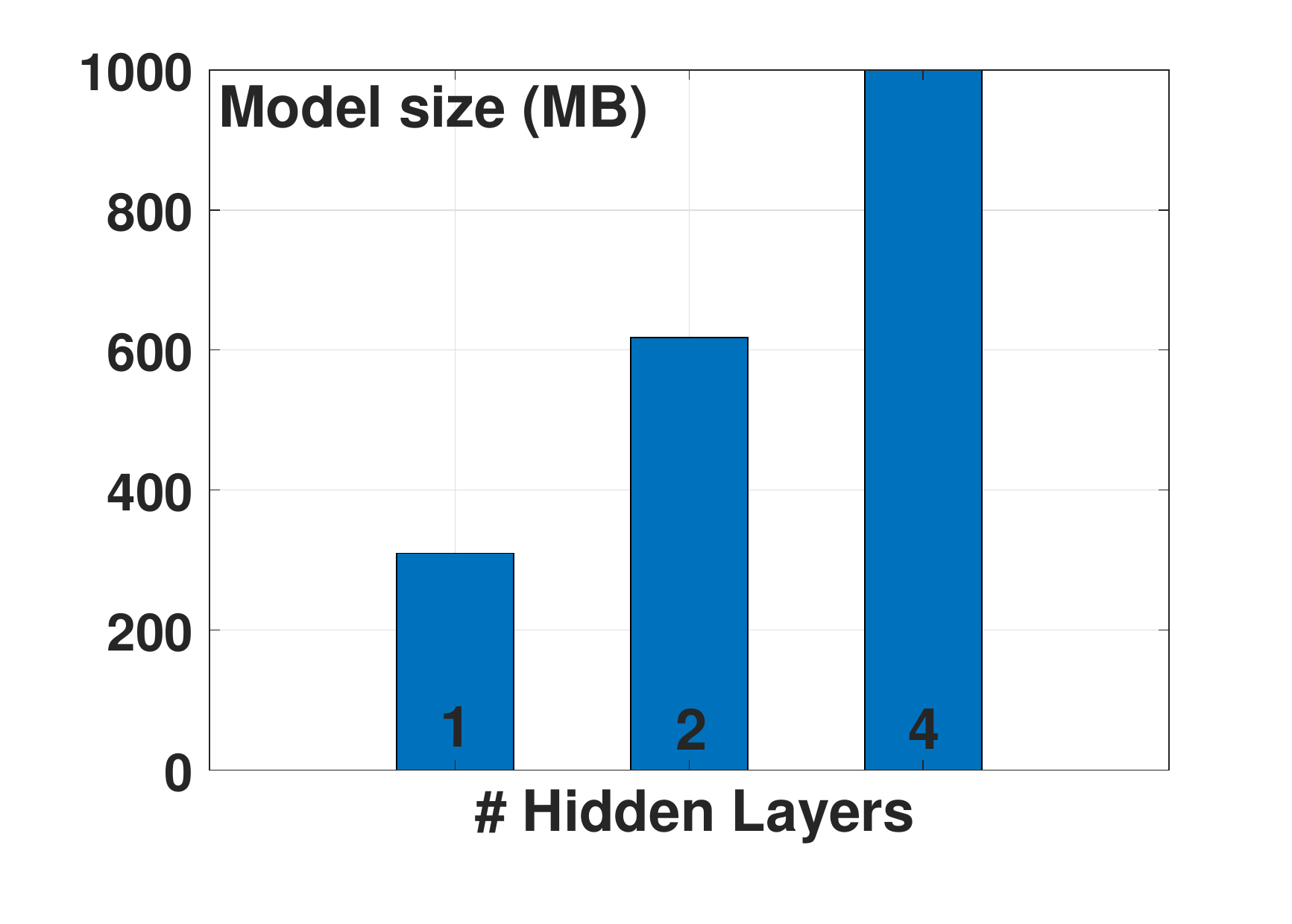}
		\caption{CNN Model Size}
		\label{fig:model_size}
	\end{subfigure}
	\vspace{-2mm}
	\caption{\rev{\small{\textbf{(a)}: Testbed: 4 RPs and a cloudlet (laptop). \textbf{(b)}: Transmit power consumption measurements and the fitted curve for the RPs. \textbf{(c)} CDF of computing cycles per task for the cloudlet. \textbf{(d)} Increasing the number of layers in CNN increases the model size (MB) up to $100\%$.}}}
	\label{fig:Tx_power}
\end{figure*}

We used  4 Raspberry Pis (RPs) as end-nodes, and a cloudlet with specs as in \cite{cloudlet}, see Fig. \ref{fig:Tx_power}a. The RPs are placed in different distances from the cloudlet, and all plots are using data from at least 50 experiments. We measured energy using a Monsoon Power Monitor, and used Python libraries and TensorFlow for the classifiers. We have used \emph{vanilla} versions of libraries and classifiers so as to facilitate observation of the results.\footnote{For instance, the memory footprint of NNs can be made smaller \cite{NN-compression, residual-learning} but such actions possibly affect their performance. Our analysis is orthogonal to such interventions.}

We measured the average power consumption when a RP transmits with different rates, Fig.~\ref{fig:Tx_power2}. Then we fitted a linear regression model that estimates the consumed power (Watts) as a function of the rate $r$, $p(r) = -0.00037r^2 + 0.0214r + 0.1277$. This result is used by OnAlgo to estimate the energy cost for each image, given the data rate in each slot (which might differ for the RPs). \rev{Also, we measured the average computing cost (cycles/task) for the classification task for a convolutional neural network (CNN) in the RPs and cloudlet. Since the images have different sizes, we observed that the computation load varies, with a mean of $441\ Mcycles$ and std. $90\ Mcycles$ for the cloudlet (see Fig.~\ref{fig:Tx_power}c), and a mean of $3044\ Mcycles$ and std. $173\ Mcycles$ for RPs. Regarding the delays, we measured device and cloudlet average processing and transmission delays and found that $D_n^{pr}=2.537$, $D_0^{pr}=0.191$ and $D_n^{tr}=0.157\ ms$. This result suggests that local processing is about 10 times slower than offloading in our system. Hence, it is possible that the extra offloading delay experienced by the devices can be worth trading off for the enhanced accuracy of the cloudlet.}

\subsubsection{Data Sets and Classifiers} We focus on image classification, a widely employed analytic task, and use two well-known data sets: \emph{(i)} \textbf{MNIST} which consists of $28\!\times\!28$ pixel handwritten digits, and includes $60$K training and $10$K test examples; \emph{(ii)} \textbf{CIFAR-10} that consists of $50$K training and $10$K test examples of $32\!\times\!32$ color images of $10$ classes. We used two very different classifiers: the normalized-distance weighted \textit{k}-nearest neighbors (KNN) algorithm~\cite{dudani1976distance}, and the more sophisticated Convolutional Neural Network (CNN), implemented with TensorFlow \cite{tensorflow}. Both classifiers output a vector where each coordinate represents the probability that the object belongs to the respective class. These classifiers differ substantially in their performance and resource requirements, hence allowing us to build diverse experiment scenarios. \rev{Our goal is to evaluate both and determine which one is more suitable depending on other system parameters like the number of available training samples at each location.}

The predictors are trained with labeled images and the outputs of the local ($d_n(s_{nt})$) and cloudlet ($d_0(s_{nt})$) classifiers. \rev{We implemented an ordinary least squares regressor and a model-free random forest that estimate $\phi_{nt}$ (dependent variables) based on the classifier outputs (independent variables).} Recall that the dependent variables are calculated using \eqref{eq:predictor-gain}. We have used training sets of different sizes and two \rev{different} regressors: \emph{(i)} a general model, where the prediction does not consider the locally inferred class as an independent variable; and \emph{(ii)} a class specific model that is based on the output of the local classifier.

\subsubsection{Benchmark Algorithms} We compare OnAlgo with \rev{three different} algorithms:
\begin{itemize}[leftmargin=4mm]
\item \emph{Accuracy-Threshold Offloading} (ATO), where a task is offloaded when the confidence of the local classifier is below a threshold, without considering the resource consumption. \rev{This is basically the non-distributed version of \cite{DDNN_1}, where if the local result is not sufficiently reliable, further CNN layers in the edge or cloud are invoked.}
\item \emph{Resource-Consumption Offloading} (RCO), where a task is offloaded when there is enough energy, without considering the expected classification improvement. 
\rev{\item \emph{Online Code Offloading and Scheduling} (OCOS)~\cite{OCOS}, where the devices always try to exploit the cloudlet's classifier, and the cloudlet tries to schedule as many tasks as possible in each slot, given its available resources.}
\end{itemize}

\subsection{Initial Measurements}

\subsubsection{Limitations of Mobile Devices} We used our testbed to verify these small resource-footprint devices require the assistance of a cloudlet. These findings are in line with previous studies, e.g., \cite{DDNN_1, DDNN_2}. The performance of a CNN model increases with the number of layers (as we will show next), but so does the model size, see Fig. \ref{fig:model_size}. We find that, even with $4$ layers, a CNN trained for CIFAR has $1$GB size and hence cannot be stored in the RPs (e.g., even more so in a smaller IoT node). Similar conclusions hold for the KNN classifier, the accuracy of which is directly linked to the number of labeled local data (KNN needs the training data available locally). Clearly, despite the efforts to reduce the size of ML models by using, for instance, compression \cite{NN-compression} or residual learning \cite{residual-learning}, the increasing complexity of analytics and the small form-factor of devices will continue to raise the local versus cloudlet execution trade off.

\begin{figure*}[t]
	\centering
	\begin{subfigure}[b]{0.3\linewidth}
		\includegraphics[scale=0.3]{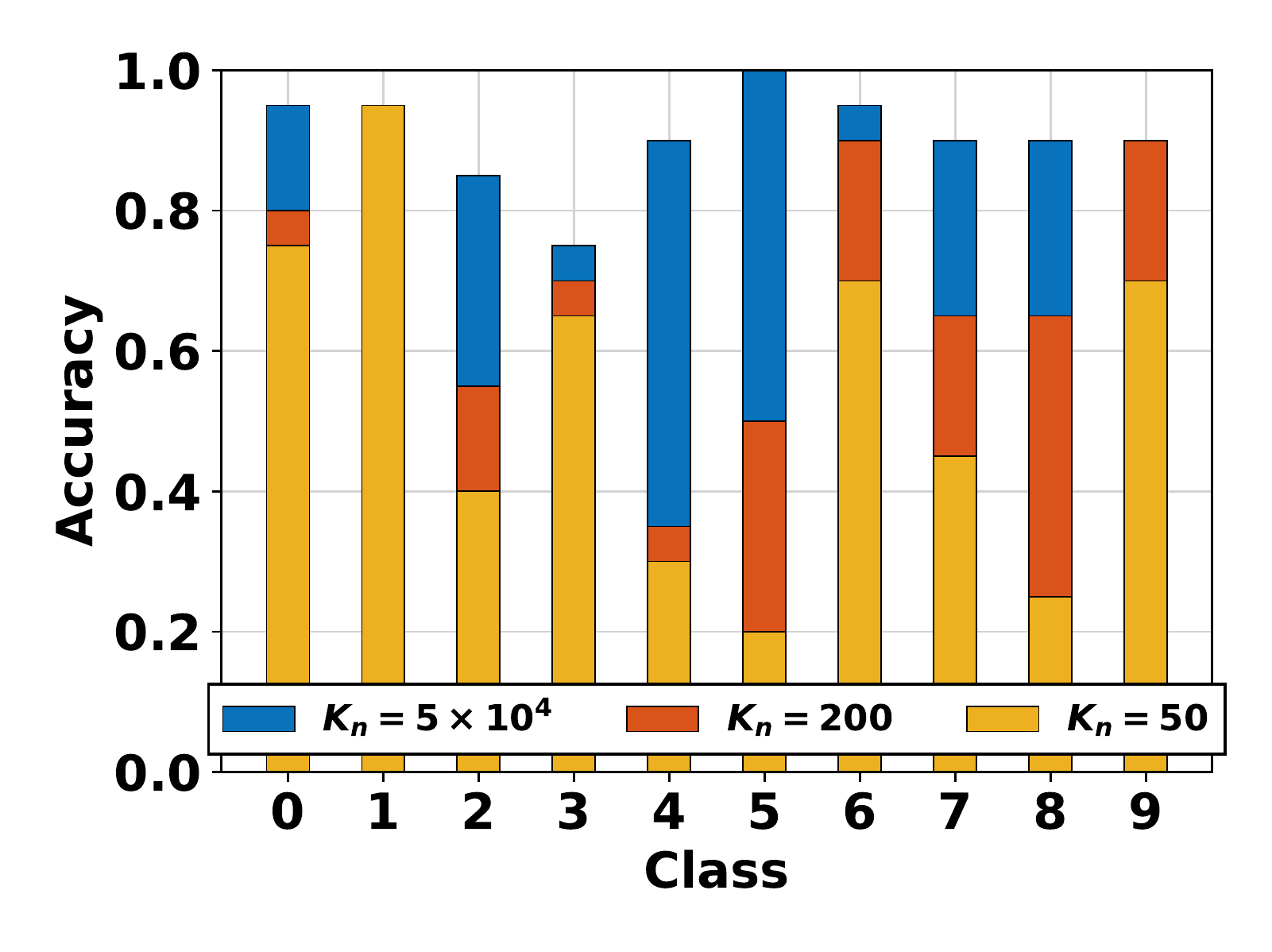}
		\caption{KNN on MNIST}
	\end{subfigure}
	~
	\begin{subfigure}[b]{0.3\linewidth}
		\includegraphics[scale=0.3]{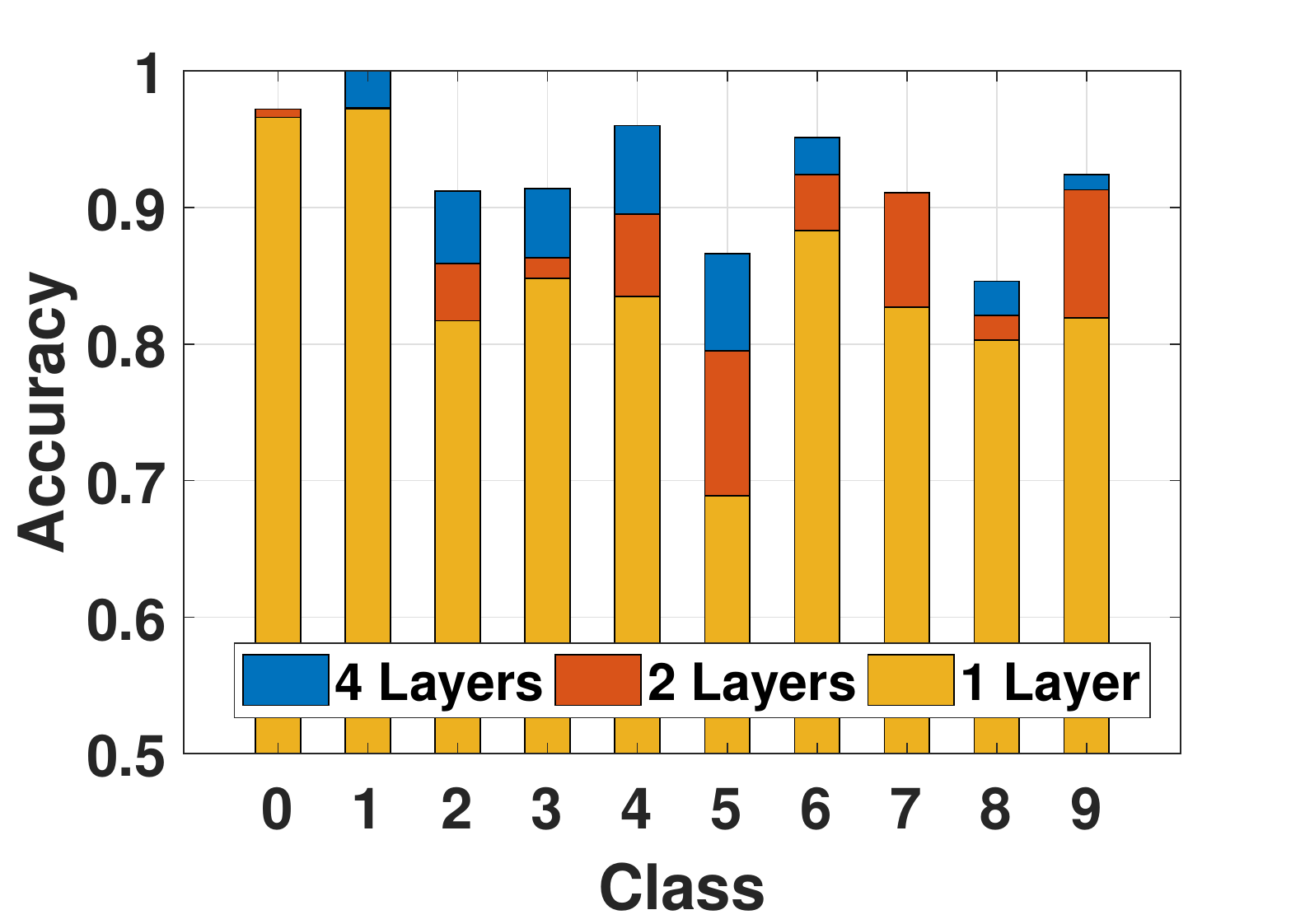}
		\caption{CNN on MNIST}
	\end{subfigure}
	~	
	\begin{subfigure}[b]{0.3\linewidth}
		\includegraphics[scale=0.323]{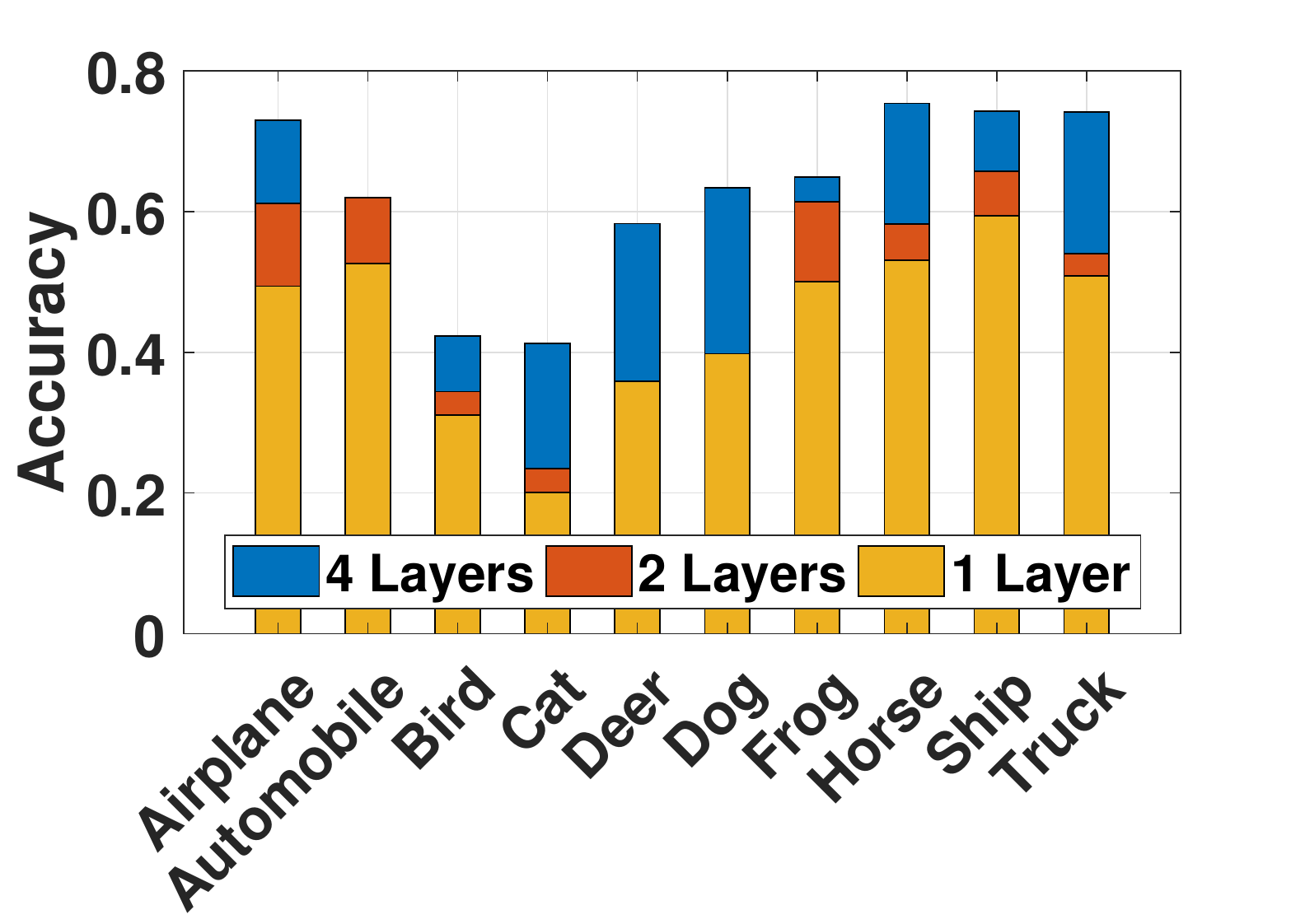}
		\caption{CNN on CIFAR}
	\end{subfigure}
	\vspace{-2mm}
	\caption{\small{Per class Accuracy of MNIST and CIFAR-10 for KNN and CNN classifiers of various labeled data sizes and hidden layers. For details of training and validation see Section V.A.}}
	\label{fig:AccuracyPerClassMNIST}
\end{figure*}

\subsubsection{Classifier and Predictor Assessment} \rev{Here we evaluate the different classifier and predictor designs towards building a more efficient system.} In Fig.~\ref{fig:AccuracyPerClassMNIST}a we see that the accuracy (defined as the ratio of correct predictions over the sum of all predictions) of the KNN classifier improves with the size $K_n$ of labeled data when applied to MNIST. Figure~\ref{fig:AccuracyPerClassMNIST}b depicts the accuracy improvement for CNN as more hidden layers are added. The performance increase is higher for the digits that are more difficult to recognize (e.g., $4$ and $5$), up to about 20\%. \rev{Notice, that the performance of the CNN classifier is superior to KNN, when we use fewer layers, or samples respectively.} In addition, we present the CNN performance on CIFAR, for $1$, $2$ and $4$ hidden layers in Fig.~\ref{fig:AccuracyPerClassMNIST}c. CIFAR is more complex than MNIST due to the properties of its objects (colored images, etc.), and this results in lower accuracy. Overall, we see that the classifier performance depends on the algorithm (KNN, CNN, etc.), the settings (datasets, layers, etc.), and differ also for each object class. \emph{Hence, an algorithm is required that can adapt to all these parameters} (as OnAlgo does). \rev{Since we have verified the superiority of CNN classifiers, we continue our evaluation using only these, instead of KNN.}

\begin{figure}[t!]
	\centering
	\begin{subfigure}[b]{0.48\linewidth}
		\centering
		\includegraphics[scale=0.3]{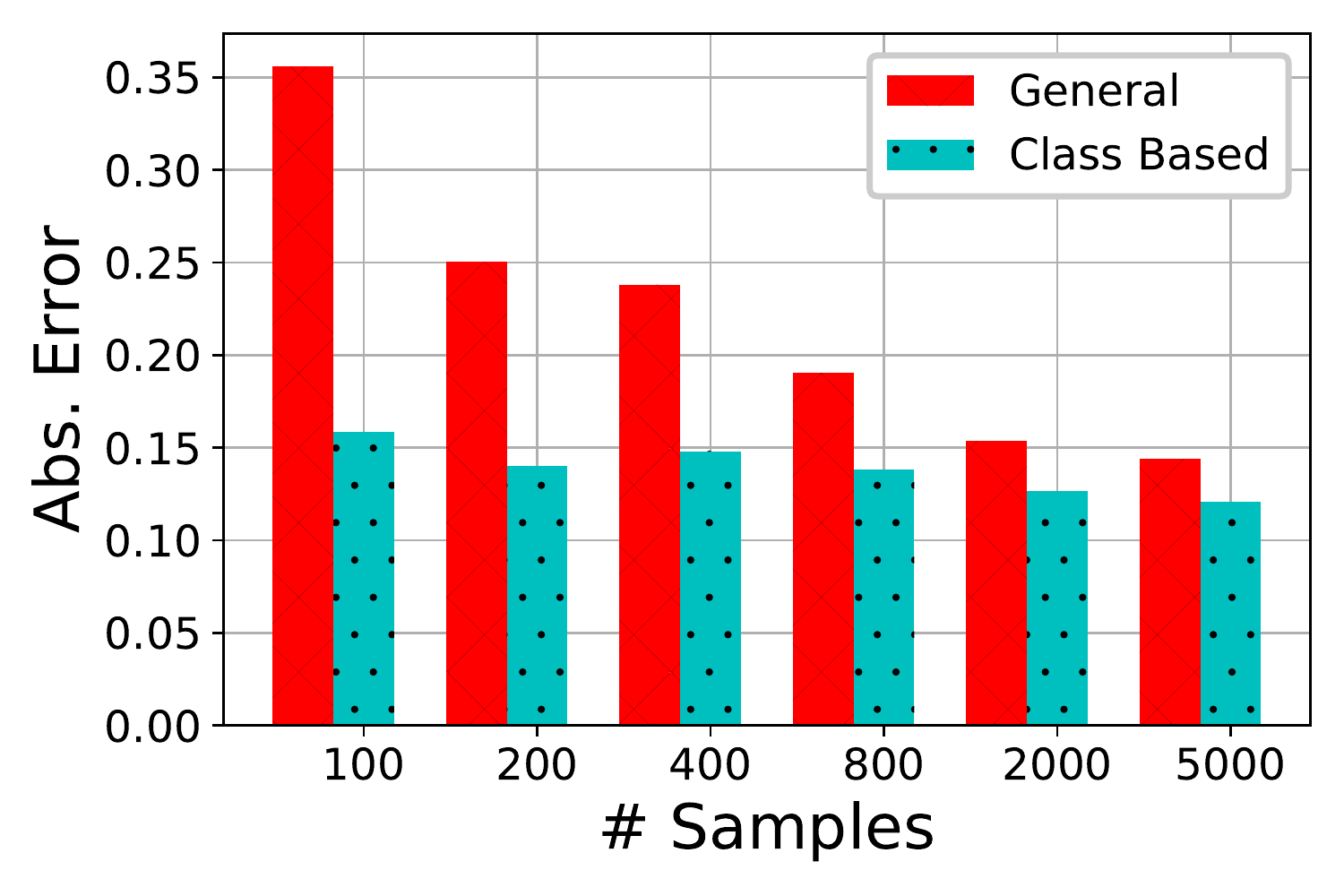}
		\caption{Linear regressor}
	\end{subfigure}
	~
	\begin{subfigure}[b]{0.48\linewidth}
		\centering
		\includegraphics[scale=0.3]{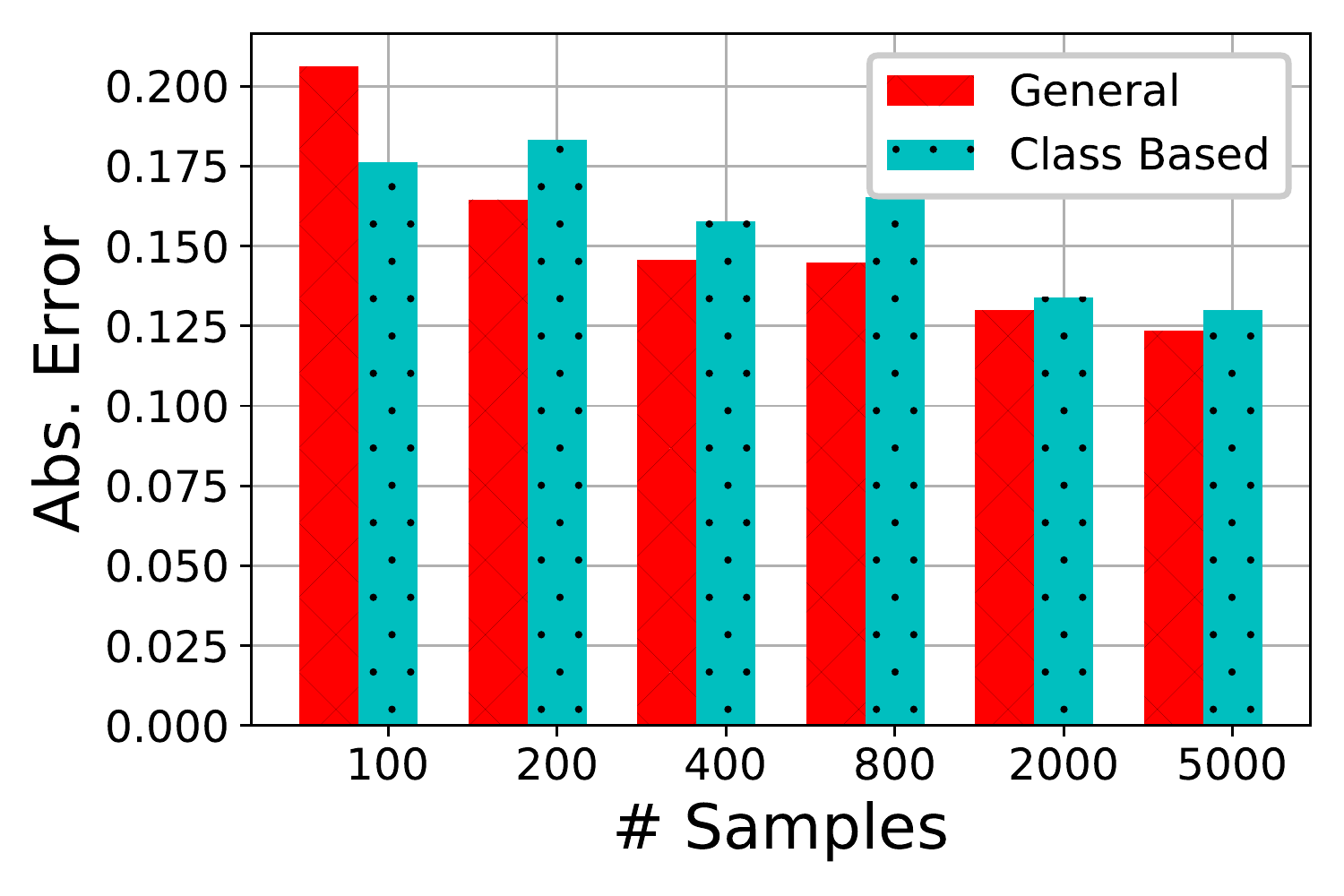}
		\caption{Random forest}
	\end{subfigure}
	\caption{Predictor assessment.}
	\label{fig:predictor}
\end{figure}

Finally, we studied the training dataset impact on the predictor's error, using both general and class-specific \rev{\textit{(i)} linear} regressors \rev{and \textit{(ii)} random forests.} In Fig.~\ref{fig:predictor}, we plot the prediction error of the accuracy improvement for both cases of general and class-specific predictors \rev{for CNN local device and cloudlet classifiers.} We observe that the random forest is superior to the simpler linear regressor only when the number of samples is small. Moreover, random forests display an inconsistency when comparing general to class-based models as the number of training samples varies. The class specific regressor for 5K samples achieves the lowest average absolute error, thus it is used throughout the following experiments, while its error is rapidly decreasing from 35\% for 100 points to 12.3\% for 5K points on the CIFAR dataset.

\subsection{Performance Evaluation}

Next, we evaluate the performance of OnAlgo in terms of achieved accuracy, offloading frequency and resource consumption. First, we evaluate OnAlgo for different values of the power consumption constraint $B_n$. \revgi{Then, we use a variable non-i.i.d. traffic load to compare its performance against the competitors, by considering these different criteria}. The traffic load is an exponentially distributed sequence of task bursts, with a uniform duration of $5-10$ seconds. This way we emulate the real-world scenario of sensor-activated cameras that generate images for short time periods.

\subsubsection{Resource Availability} 
\begin{figure}[t!]
	\centering
	\begin{subfigure}[b]{0.48\linewidth}
		\centering
		\includegraphics[scale=0.29]{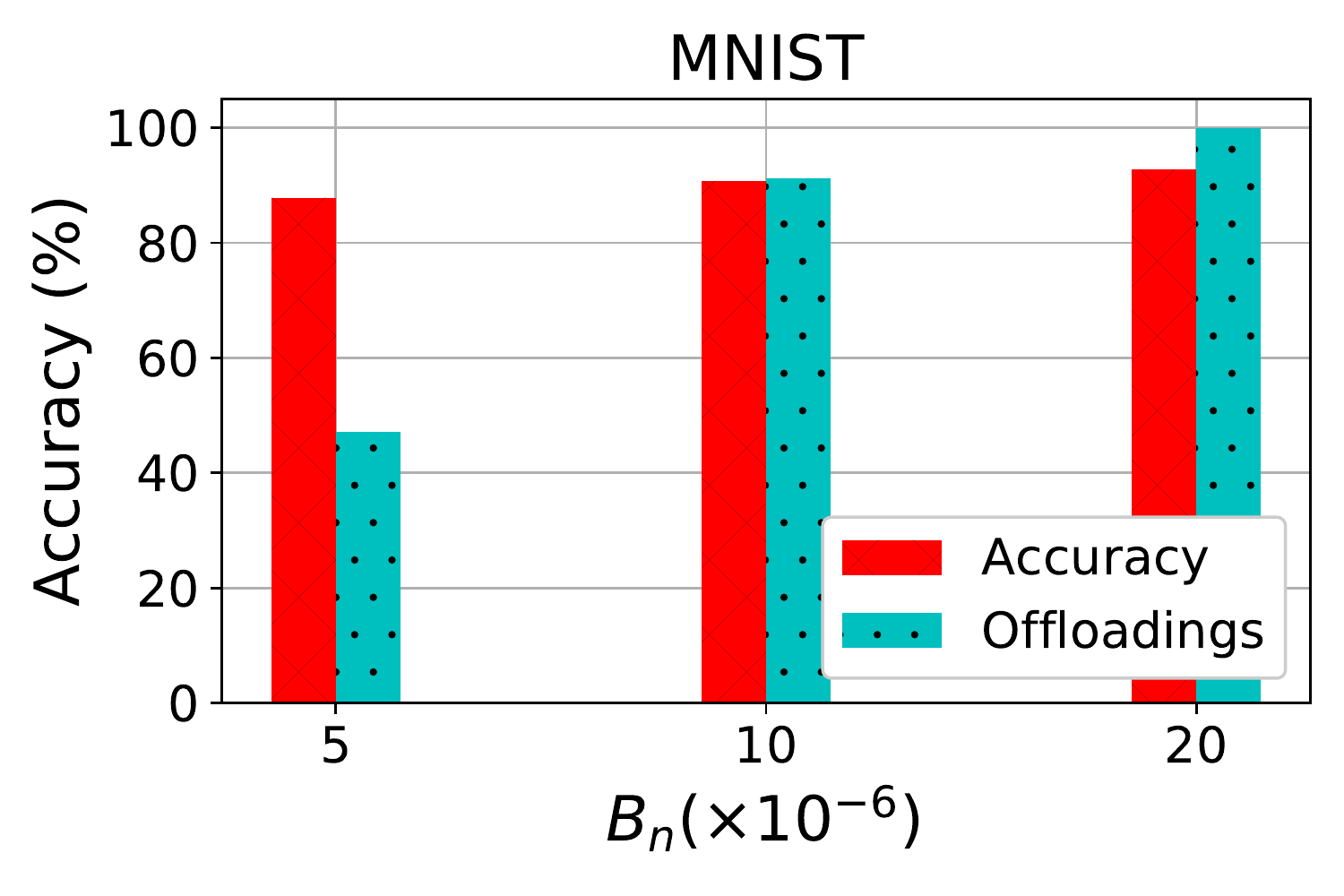}
		\caption{}
	\end{subfigure}
	\begin{subfigure}[b]{0.48\linewidth}
		\centering
		\includegraphics[scale=0.29]{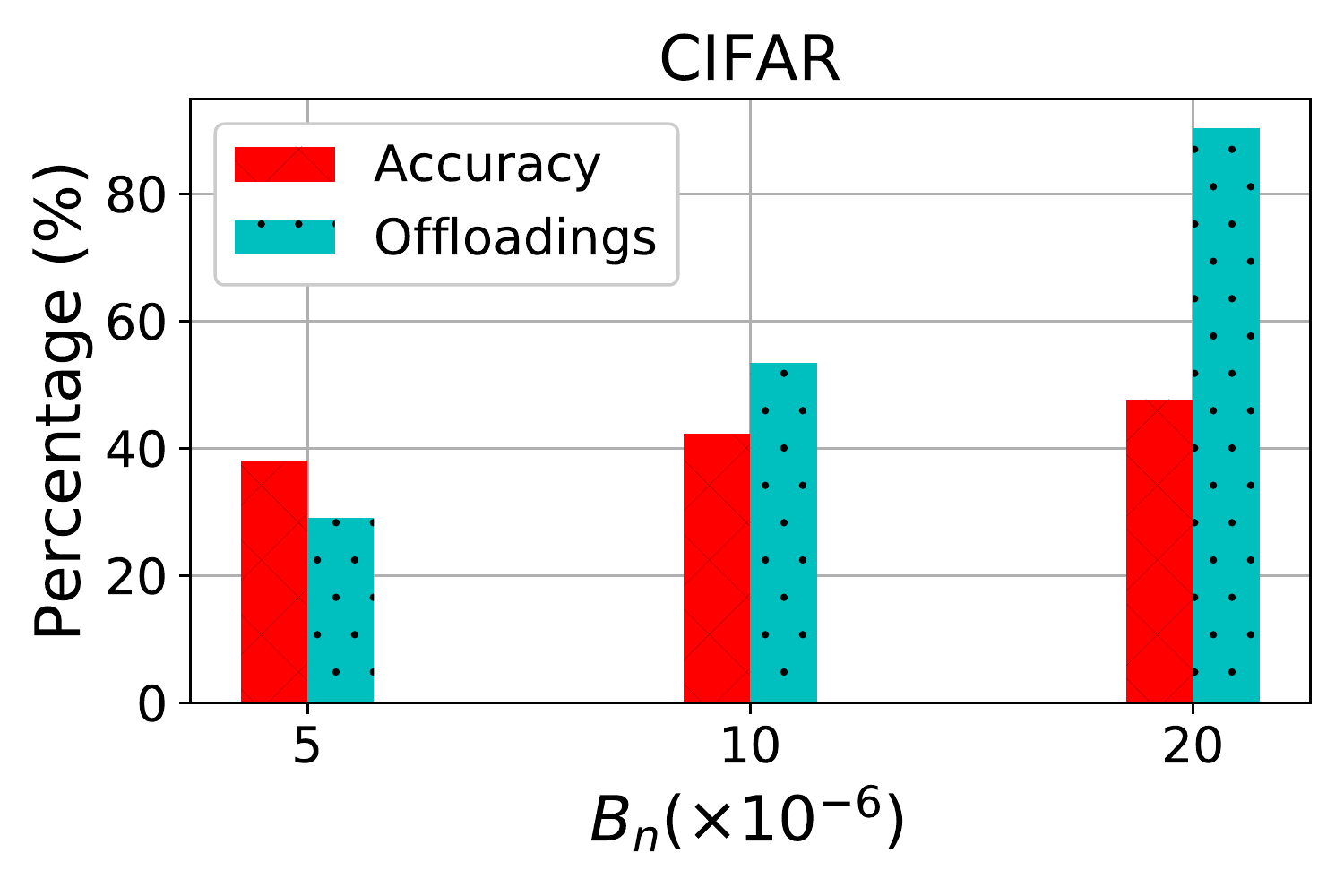}
		\caption{}
	\end{subfigure}
	\caption{\small{Accuracy and offloading percentage of OnAlgo for various resource constraints, on MNIST and CIFAR-10.}}
	\label{fig:Resource_results}
\end{figure}

We evaluate OnAlgo, by using a 1-layer CNN for the RPs and a 4-layer CNN for the cloudlet. In Fig. \ref{fig:Resource_results} we show the average accuracy achieved by the four devices, as well as the fraction of requests offloaded to the cloudlet when we vary the devices' power budget $B_n$, for MNIST and CIFAR. Evidently, as $B_n$ increases there are more opportunities for exploiting the cloudlet and obtaining a better result than the local classifier. Furthermore, some interesting remarks can be made by comparing the two datasets. As shown in Fig. \ref{fig:AccuracyPerClassMNIST}(b-c), MNIST is easier to classify and the gain of using a better classifier is not as important as on the CIFAR dataset. In particular, with MNIST the gains are about $6\%$ in accuracy as the resources (and thus the offloaded tasks) increase. With CIFAR, on the other hand, the potential performance gain when using the cloudlet is higher; and as $B_n$ increases, the accuracy gains are up to $15\%$. These two experiments demonstrate \emph{the agility of our algorithm, which assesses the potential accuracy gains and shapes accordingly the offloading strategy, based on resource availability.}  

\subsubsection{Comparison with Benchmarks} 
\begin{figure*}[t!]
	\centering
	\begin{subfigure}[b]{0.24\linewidth}
		\centering
		\includegraphics[scale=0.3]{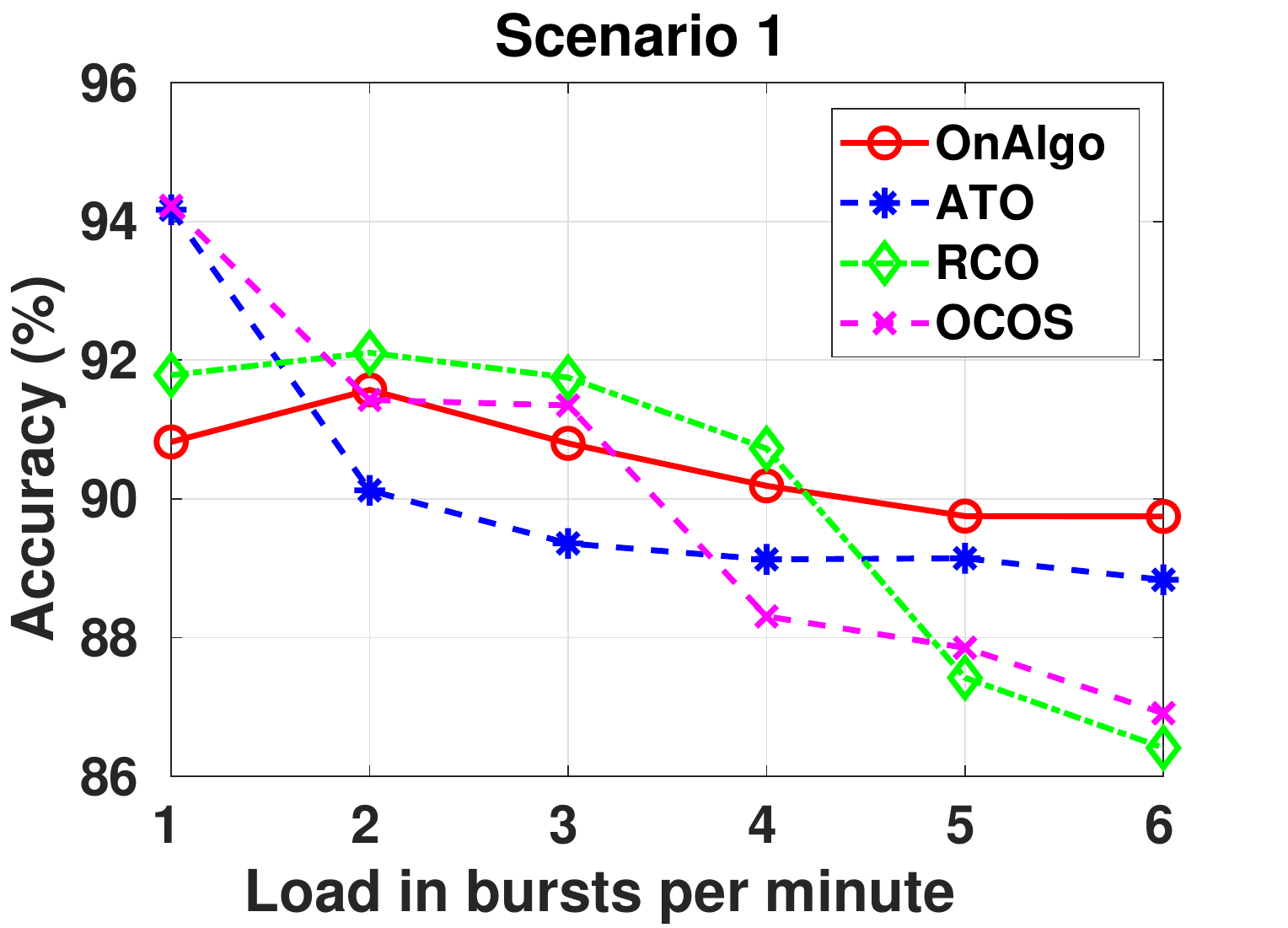}
		\caption{}
	\end{subfigure}
	\begin{subfigure}[b]{0.24\linewidth}
		\centering
		\includegraphics[scale=0.3]{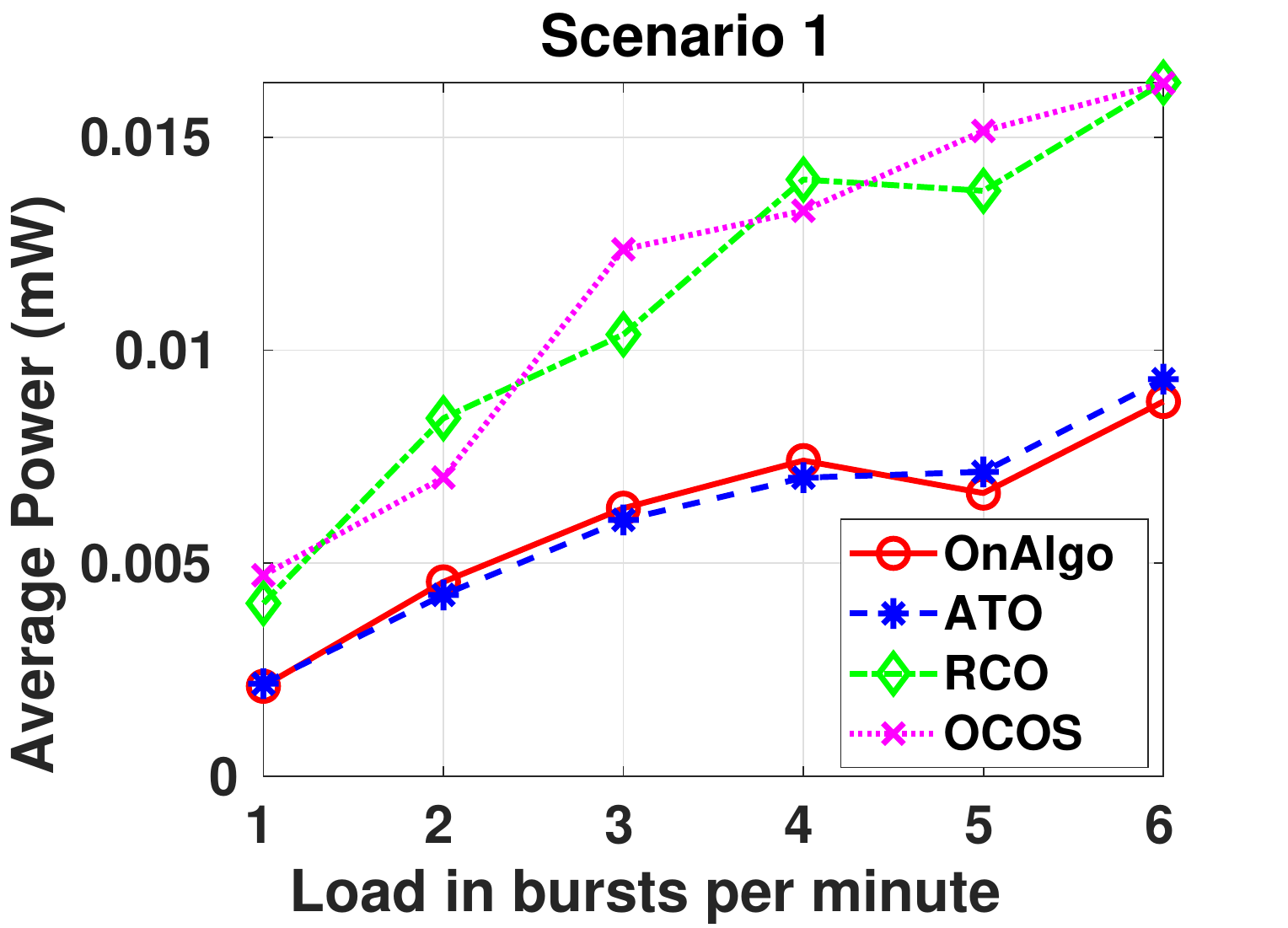}
		\caption{}
	\end{subfigure}
	\begin{subfigure}[b]{0.24\linewidth}
		\centering
		\includegraphics[scale=0.3]{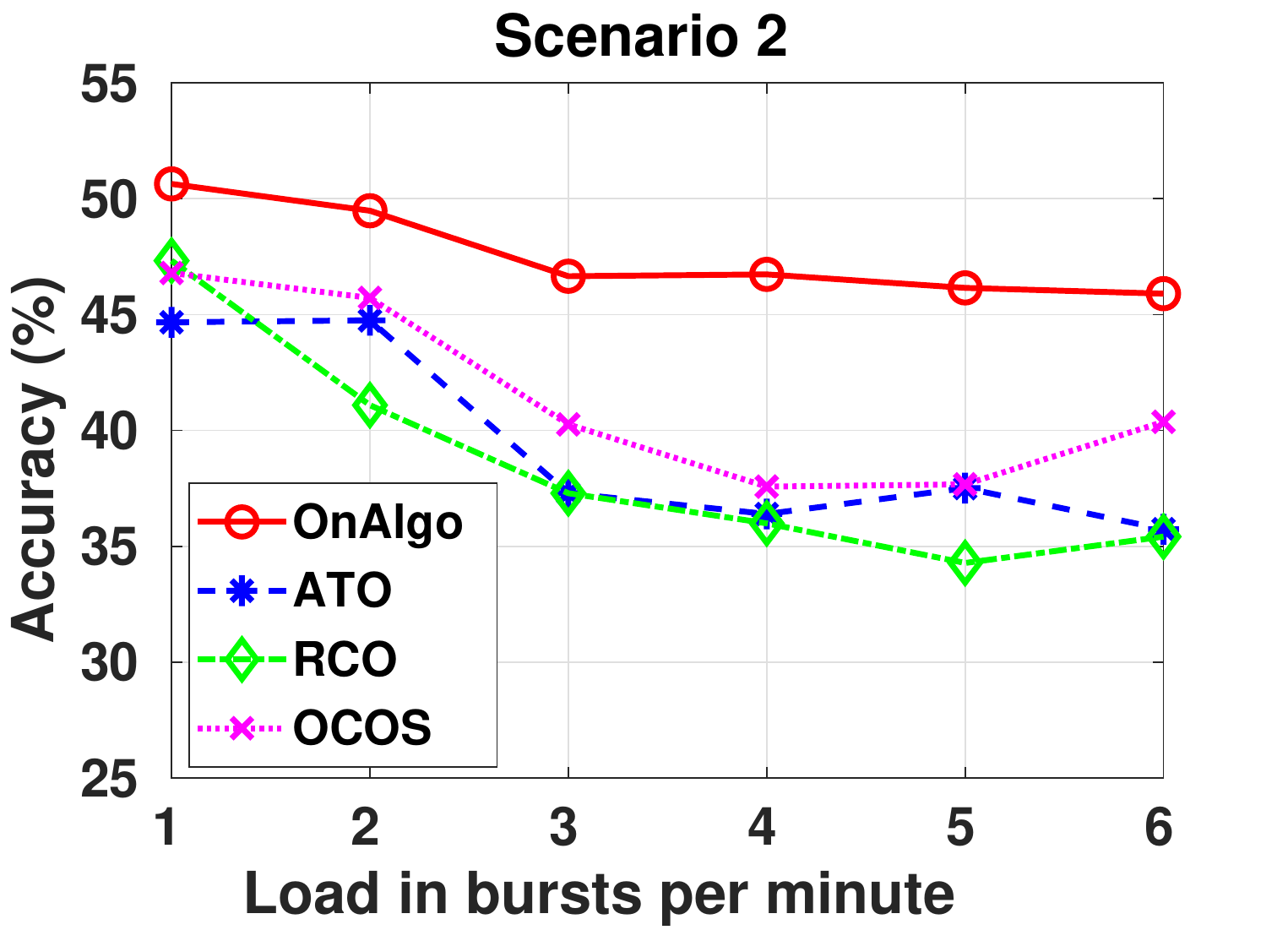}
		\caption{}
	\end{subfigure}
	\begin{subfigure}[b]{0.24\linewidth}
		\centering
		\includegraphics[scale=0.3]{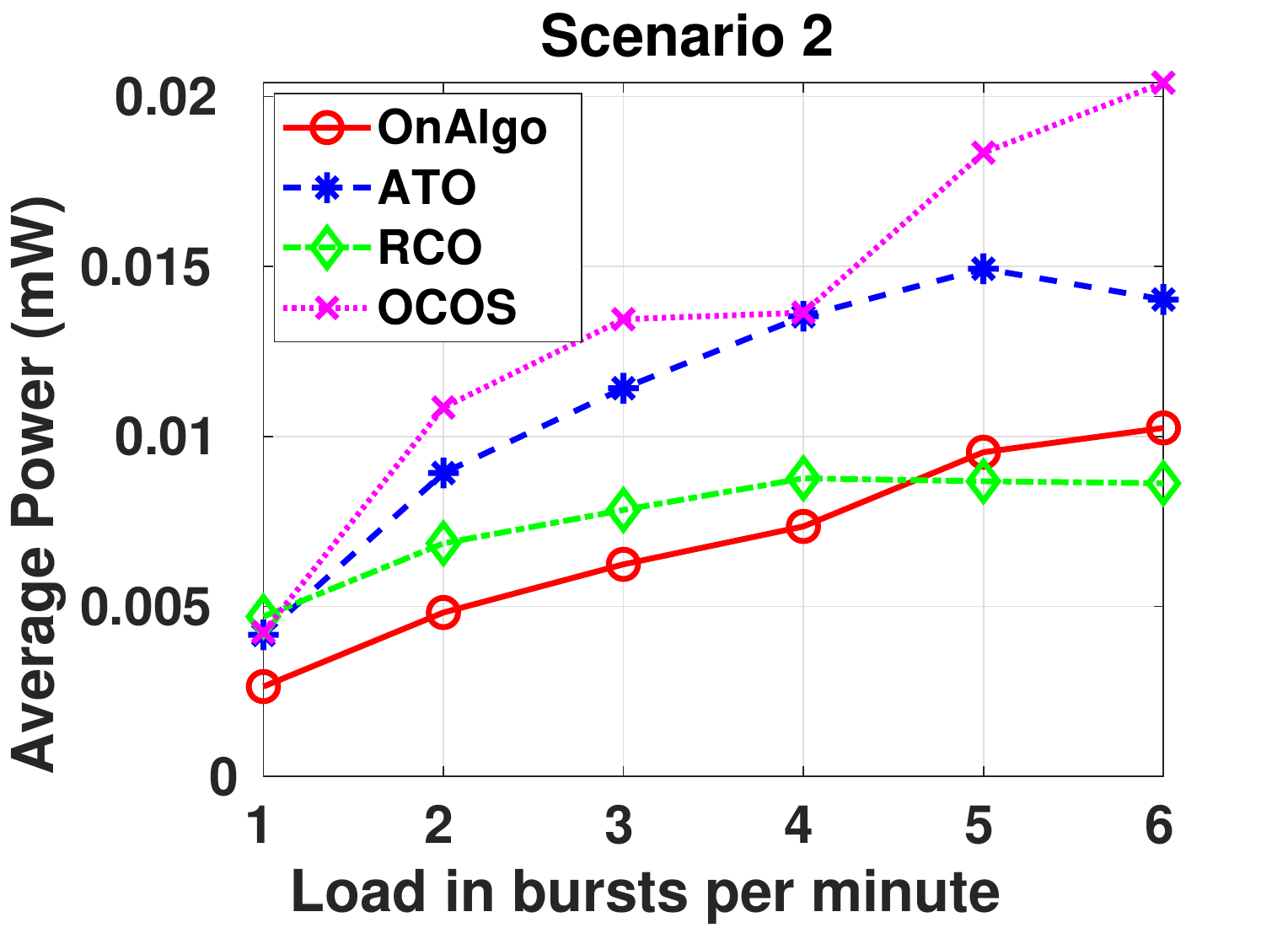}
		\caption{}
	\end{subfigure}
	\caption{\small{\revgi{Comparison of different offloading algorithms w.r.t. their accuracy and energy cost, under different task load conditions.}}}
	\label{fig:comparison_1}
\end{figure*}

Next, we compare OnAlgo to \rev{ATO, RCO and OCOS for a varying non-i.i.d. traffic load in Fig.~\ref{fig:comparison_1}} \revgi{using the criterion of accuracy and power consumption. Ideally, we would like an algorithm to perform well in both these dimensions.} To ensure a realistic comparison, we set the rule for all algorithms that the cloudlet will not serve any task if the computing capacity constraint is violated; while for RCO the availability of energy is determined by computing the average consumption by each device during the experiment. We employ two scenarios to demonstrate \revgi{the algorithms' performance and energy costs} under different data sets and resource availability states.

\textbf{Scenario 1:} \textit{Low accuracy improvement; high resources}. In this case, we set\footnote{We have explicitly set a small power budget so as to highlight the impact of power constraints on the system performance; higher power budgets will still be a bottleneck for higher task request rates or images of larger size.} $B_n\!=\!0.02\ mW$, $H\!=\!2\ GHz$ allowing the devices to offload many tasks, and the MNIST dataset (has small improvement between 1 layer and 4 layer CNNs). We depict the average accuracy achieved by the devices and the average power consumption versus the task load in bursts per minute in Fig. \ref{fig:comparison_1}a and \ref{fig:comparison_1}b respectively. We observe that OnAlgo shows a smaller slope in the decrease of accuracy, as the load increases than all the competitors. The performance of ATO quickly drops because the cloudlet's resources are insufficient for high loads. \revgi{RCO's performance is good for the most part, but it quickly deteriorates for high task loads as the devices refrain from offloading due to the power constraints. It is interesting also to note that RCO even outperforms OnAlgo in terms of accuracy (by approx. $2\%$) but this happens at the expense of larger energy cost, namely it spends more than double the energy of OnAlgo.} OCOS performs similarly to RCO since performance degradation is caused by cloudlet resource exhaustion. The problem with both algorithms is that they do not offload intelligently, based on both the improvement potential \emph{and} the availability of resources. \revgi{Hence, when considering both performance and energy cost criteria, and especially in the non-trivial higher load cases they are significantly outperformed by OnAlgo.}

\textbf{Scenario 2:} \textit{High accuracy improvement; low resources}. The settings for this scenario are $B_n = 0.01\ mW$, $H = 500\ MHz$ not allowing many offloadings and cloudlet classifications. We used the CIFAR dataset that demonstrates a substantial performance difference between local and cloudlet classifiers. We see from Fig.~\ref{fig:comparison_1}c that OnAlgo is up to 12\% more accurate than ATO/RCO for high task load, and in any case significantly higher than in Scenario 1. OCOS performs slightly better than ATO/RCO, but at the cost of very high power consumption. Since the potential of improvement is higher in Scenario 2, ATO marginally outperforms RCO by spending up to 50\% more power than RCO (see Fig.~\ref{fig:comparison_1}d). OnAlgo consumes about 50 \% less power than OCOS since the latter always tries to offloads tasks but does not leverage the cloudlet efficiently due to the lack of computing capacity.

Summing up the 2 scenarios above, we see that OnAlgo achieves a smooth performance across varying traffic loads, while its competitors struggle, especially as the load increases. Moreover, it achieves reasonable power consumption regardless of the resource availability as opposed to RCO in Scenario 1, ATO in Scenario 2, and OCOS in both scenarios. \revgi{Even when in some cases OnAlgo is being outperformed by some competitor with respect to one criterion (e.g., by RCO w.r.t. accuracy in Scenario 1), this happens at the expense of losing at much larger rates w.r.t. the other criterion (power consumption).}

\subsubsection{Trade-off Analysis} 
\begin{figure}[t!]
	\centering
	\begin{subfigure}[b]{0.48\linewidth}
		\centering
		\includegraphics[scale=0.41,trim=2cm 0cm 2cm 0cm, clip=true]{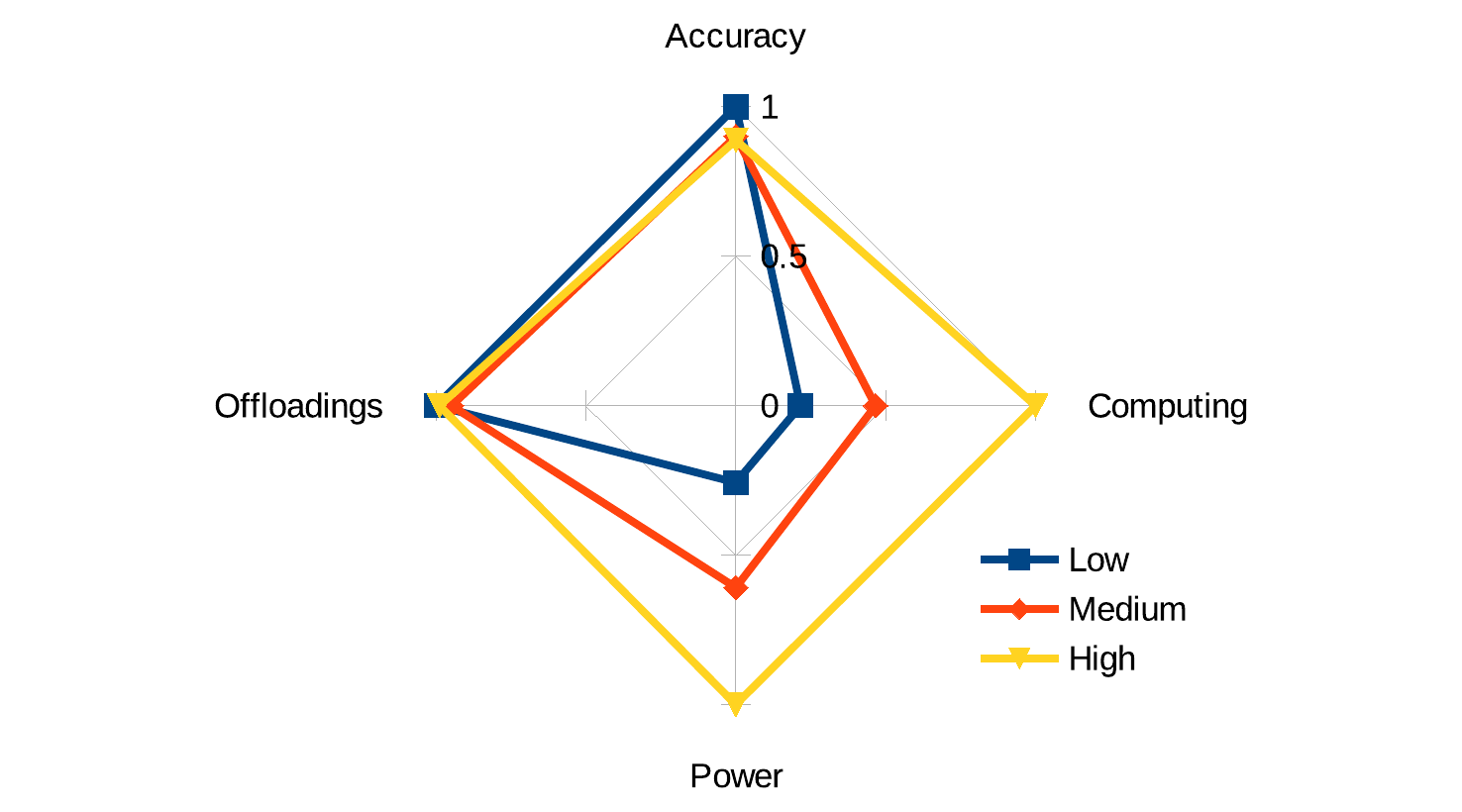}
		\caption{}
	\end{subfigure}
	~
	\begin{subfigure}[b]{0.48\linewidth}
		\centering
		\includegraphics[scale=0.41,trim=1.2cm 0cm 1.2cm 0cm, clip=true]{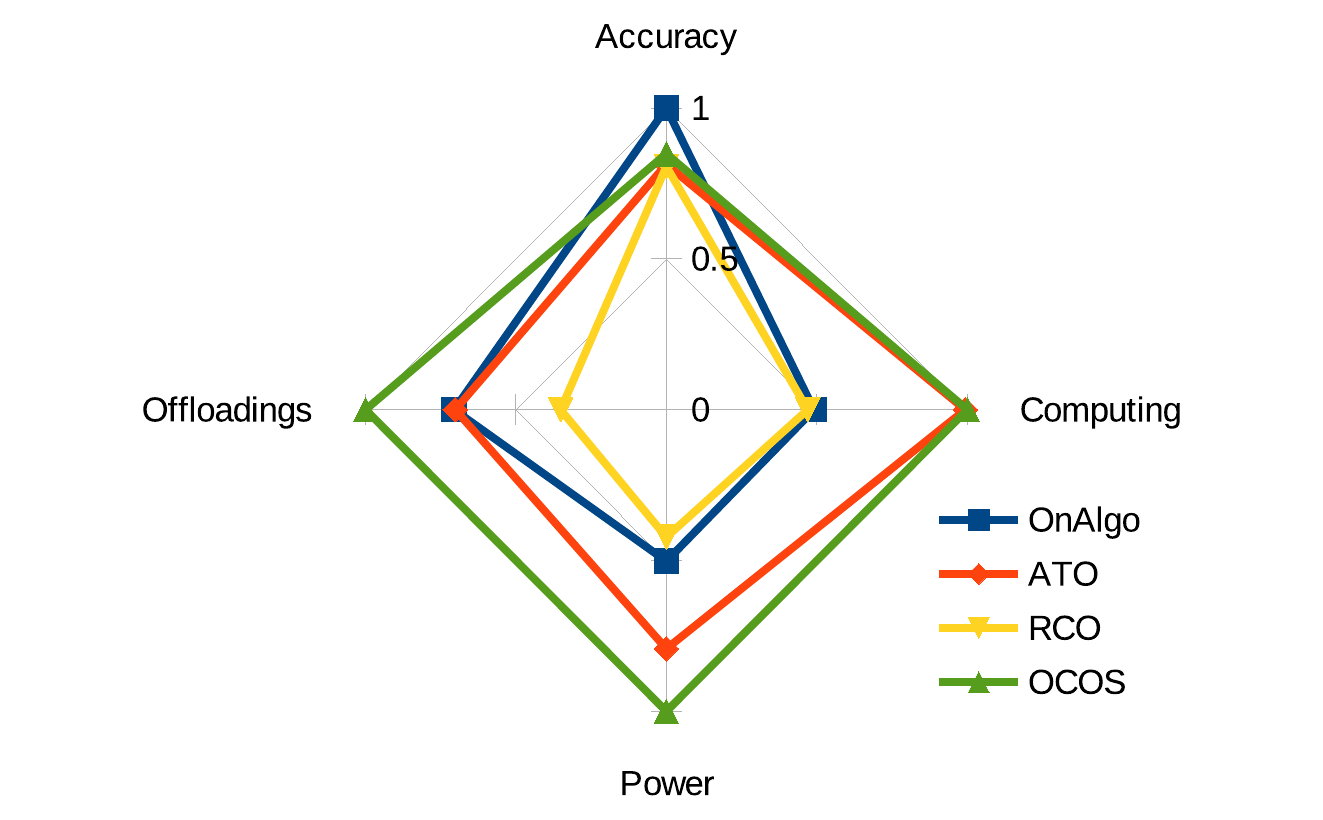}
		\caption{}
	\end{subfigure}
\vspace{-3mm}
	\caption{\small{\rev{Comparison of different key metrics (normalized): (a) OnAlgo for low, medium and high traffic load. (b) Algorithm comparison for high load in scenario 2.}}}
	\label{fig:tradeoffs}
\end{figure}

Next we demonstrate the trade-offs between number of offloadings, accuracy and resource consumption between OnAlgo and the competitor algorithms using net graphs. Fig.~\ref{fig:tradeoffs}a displays the performance of OnAlgo for low medium and high task load. Observe that as the load increases, OnAlgo rapidly increases resource consumption to maintain high accuracy. For instance, comparing low to high load, we see that performance drops only by about 7\% as the computing and power consumption is increased by 75\%. In Fig.~\ref{fig:tradeoffs}b we compare the same metrics for high traffic load, and the different competitors. Observe that OnAlgo achieves the highest accuracy, while being (closely) second best (behind RCO) in terms of computing resource and power consumption. Moreover it achieves high accuracy despite offloading less frequently than OCOS, due to the intelligent way it makes the offloading decisions. In summary, OnAlgo achieves the \emph{highest} accuracy between the competitors, and at the same time has a \emph{moderate} resource consumption.

\begin{figure}[t!]
	\begin{subfigure}[b]{0.48\linewidth}
		\centering
		\includegraphics[scale=0.37,trim=1.5cm 0cm 1.5cm 0cm, clip=true]{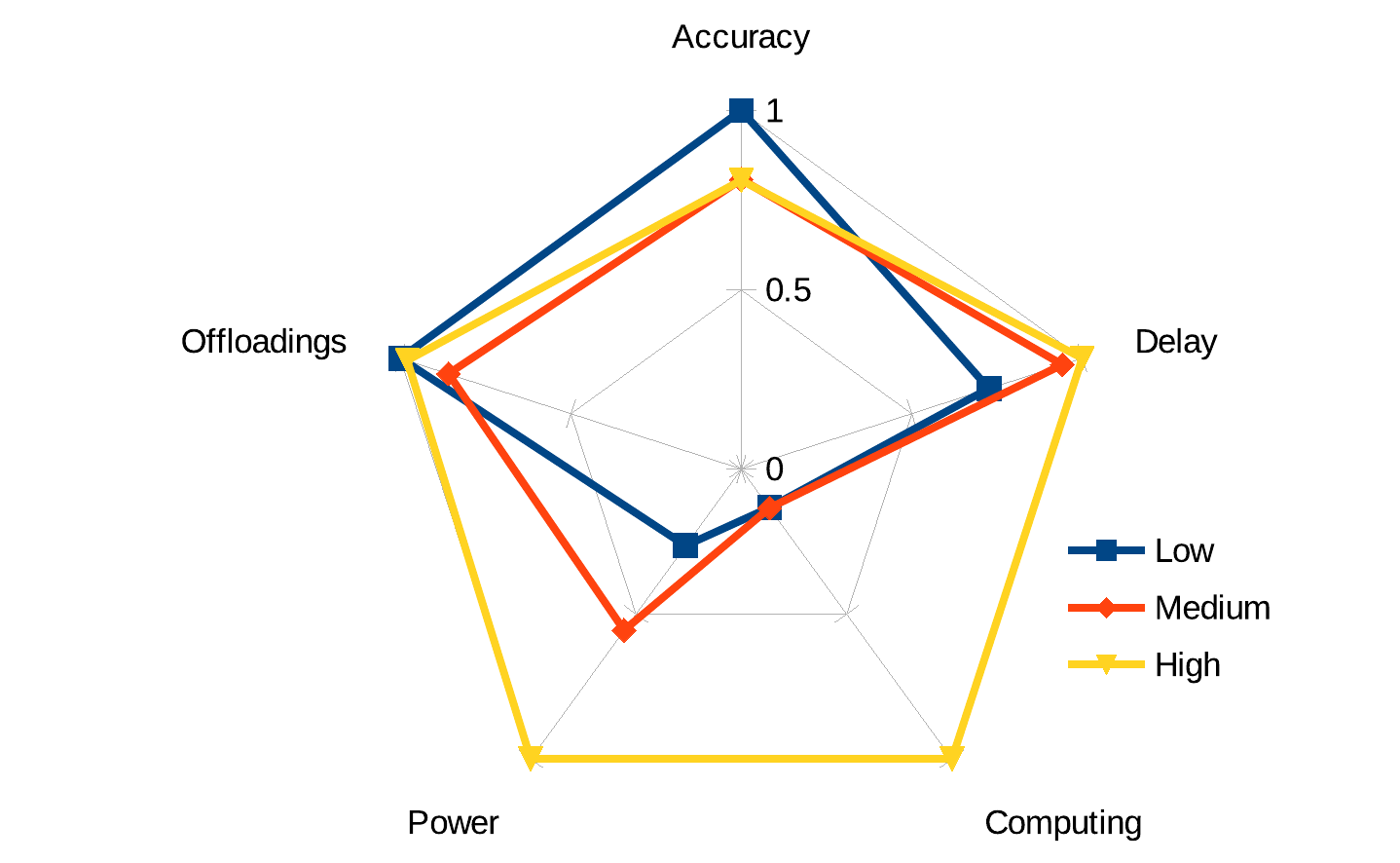}
		\caption{}
	\end{subfigure}
	~
	\begin{subfigure}[b]{0.48\linewidth}
		\centering
		\includegraphics[scale=0.3]{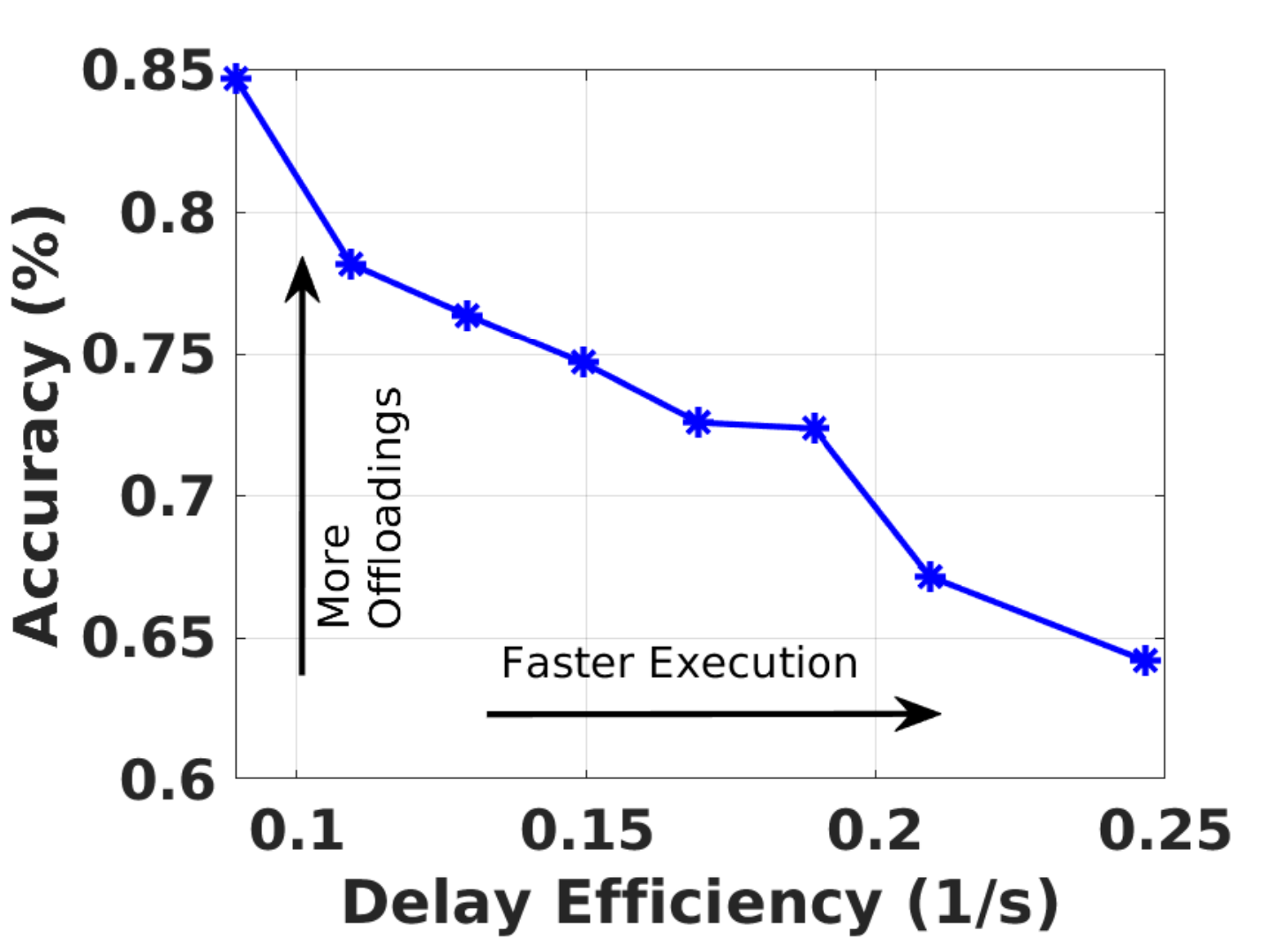}
		\caption{}
	\end{subfigure}
\vspace{-3mm}
	\caption{\small{(a) OnAlgo performance for problem (P3). (b) Pareto front between accuracy and delay efficiency obtained by tuning $\zeta$.}}
	\label{fig:tradeoffs-delay}
\end{figure}

Next, in Fig.~\ref{fig:tradeoffs-delay} we explore the accuracy-resource consumption-delay trade-off when problem (P3) is considered, i.e. total delay is jointly optimized with accuracy. Notice in Fig.~\ref{fig:tradeoffs-delay}a, that the increasing traffic load will not only result in lower accuracy (about 20\%) and higher resource consumption, but also in significantly higher delay (up to 25\%). \revv{Hence, despite consuming extra resources in high load cases, OnAlgo still maintains high accuracy standards.} Finally, Fig.~\ref{fig:tradeoffs-delay}b displays the Pareto front between accuracy and delay\footnote{In fact delay is inversed (1/s) so that increasing the value towards either the x-axis or the y-axis yields better performance with respect to the relevant metric.}. This shows the effect of parameter $\zeta$ \revv{(ranging from 0.1 to 0.3)} on the resulted offloading policy and consequently on the performance of accuracy and delay. For instance, in order to \revv{double the delay efficiency (from 0.1 to 0.2)}, we would have to sacrifice roughly 10\% accuracy, by offloading less frequently.

\section{Related Work} \label{sec:related}
\textbf{Edge \& Distributed Computing}. Most solutions partition compute-intense mobile applications and offload them to cloud~\cite{CloneCloud}. This approach cannot support applications with stringent requirements due to possible large delays in data transfers \cite{ha-cloud-delay}. Cloudlets on the other hand, achieve lower delay by leveraging edge computing~\cite{Cloudlets, Smart_cities} but have limited serving capacity. \revgi{A different line of work proposes the distribution of tasks among collaborating nodes \cite{dos-anjos, hai-iot2021, ghosh} using intuitive allocation metrics or static optimization models.} Hence, there is need for an intelligent cloudlet offloading strategy and this idea lies at the core of our proposal which, unlike previous works: \emph{(i)} considers the quality of outcome (accuracy) and resource costs of devices and cloudlet; and \emph{(ii)} is adaptive and oblivious to statistics of system parameters and user requests. O

Previous works in this area consider simple performance criteria, such as reducing the computation load and focus on the architecture design. For example, Misco~\cite{Misco} and CWC~\cite{CWC} implement frameworks for parallel task execution on mobile devices; and similarly MobiStreams~\cite{MobiStreams}, Swing~\cite{Swing} and~\cite{OKeefe15networkaware}, focus on collaborative data stream computations. These systems either do not optimize the offloading policy~\cite{MobiStreams} or use heuristics that do not cater for task accuracy~\cite{Swing},~\cite{OKeefe15networkaware}. Instead, OnAlgo ensures optimal performance, subject to resource availability, even when the latter is unknown.

\textbf{System Designs for Mobile Analytics}. The increasing importance of these services has motivated the design of wireless systems that can execute such tasks. \revgi{For instance, \cite{vrarDeep,lane,MobileNets} tailor deep neural networks for execution in mobile devices, and \cite{wang-infocom21} focuses on how to maximize accuracy in edge-cloud deployments. These works however, focus either only on execution delay or accuracy.} Glimpse~\cite{glimpse} reduces delay in video tracking applications using an active cache of frames at the device; Cachier~\cite{cachier} uses edge servers as caches for image recognition requests so as to minimize latency; and Precog~\cite{precog-sec17} prefetches trained classifiers on devices to accelerate image recognition. In a different approach, \cite{mcdnn} selects in runtime the the DNN size, in order to balance accuracy and resource consumption. Similarly, \cite{deepdecision} considers a richer set of decisions, including model selection; image compression; and frame rate, aiming to maximize the accuracy of frames. Finally, \cite{mobiqor} minimizes execution time and energy cost for a single device, for known system parameters and task loads; while \cite{hetero-edge} optimizes again delay but through the orchestration of the edge resources. The plethora of such system proposals underlines the necessity for an \emph{analytical framework} for task outsourcing that can optimize performance.

\textbf{Optimization of Analytics}. Prior analytical works in the context of computation offloading focus on different metrics, such as the number of served requests, e.g., see \cite{xu-ToN} and overview in \cite{letaief-edge-tutorial}, and hence are not applicable here. In our previous work \cite{TCCN}, we proposed a \emph{static} optimization framework for a peer-to-peer collaborative task execution scheme, which does not employ predictions of gains nor accounts for computation constraints. \revgi{The authors of \cite{zhang_2021} employ a Lyapunov optimization approach to configure a video analytics application towards balancing the accuracy and energy costs, under i.i.d. requests and system dynamics}. In \cite{Yang_2020}, video quality and computing resources are selected to maximize the approximate analytics accuracy. \revgi{FastVA \cite{FastVA} is a video analytics system that leverages neural processing units at the mobile devices and proposes a heuristic offloading policy towards maximizing accuracy.} Other works that cater for accuracy either rely on heuristics or static models and complete knowledge of system parameters \cite{deepdecision, mcdnn, yi-SEC17}. 

Clearly, these assumptions are invalid for many practical systems where the expected accuracy improvements, power availability, wireless channels, and cloudlet resources not only vary with time, but often do not follow an i.i.d. process. This renders the application of max-weight type of policies \cite{tassiulas-book} inefficient. Our approach is fundamentally different and leads to a more robust algorithm that converges as long as the perturbations are bounded (in each slot), and have well defined mean values (which can be unknown). Our methodology is inspired by dual averaging and primal recovery algorithms for static problems, see \cite{lindberg, nedic-subgrad-siam}, \cite{victor-2}. We have extended here this idea and succeeded in obtaining deterministic bounds and for a broad range of perturbations. It is also important that the employed algorithm is lightweight and amenable to distributed execution, hence can be implemented as a network protocol. \revgi{This is in contrast with other optimization approaches, e.g. using Bayesian learning \cite{automl, edgebol}, which require a centralized computation-demanding execution.}

\textbf{Improvement of ML Models}. Clearly, despite the efforts to improve the execution of analytics at small devices, e.g., by compressing NN models \cite{NN-compression} or using residual learning \cite{residual-learning}, the trade off between low-accuracy local and high-accuracy cloudlet execution is still important due to the increasing number and complexity of these tasks. This observation has spurred efforts for designing fast multi-tier deep neural networks \cite{DDNN_1}; for dynamic model selection \cite{dnn-dynamic, mcdnn}; and for threshold-based task allocation to DNNs \cite{DDNN_2}; see also discussion in \cite{deepdecision}. These works are orthogonal to our approach and can be readily incorporated in our framework.

\section{Conclusions} \label{sec:conclusions}
We propose the idea of augmenting the execution of data analytics at end devices with more accurate libraries, or routines, running at a cloudlet. The key feature of our proposal is a dynamic and distributed algorithm that makes the outsourcing decisions based on the expected performance improvement, and the available resources at the devices and cloudlet. It was shown, theoretically and through experiments, that this \emph{joint performance-costs} design outperforms other efforts that do not cater for the analytics accuracy or the resource availability. The proposed algorithm achieves near-optimal performance in a deterministic fashion, and under minimal assumptions about the system behavior. Namely it suffices the perturbations to be bounded in each slot and have well-defined means. \revgi{This makes it ideal for the problem at hand where the stochastic effects (e.g., expected accuracy gains) might not follow an i.i.d. or a Markov-modulated process.}


\section{Appendix} \label{sec:appendix}

\revgi{
\textbf{Proof of Lemma \ref{lem:complower}:} For any $\bm \theta \in \mathbb R^{N+1}_+$ we can write:
\begin{align}
\Vert \bm \lambda_{t+1} - \bm \theta \Vert^2\!&=\!\Vert [\bm \lambda_t\!+\!\alpha_t g_t(\bm y_t) ]^+\!-\!\bm \theta \Vert^2 \le \Vert \bm \lambda_t\!+\!\alpha_t g_t(\bm y_t)\!-\!\bm \theta \Vert^2 \notag \\
&=\!\Vert\bm  \lambda_t\!-\!\bm \theta \Vert^2\!+\!\alpha_t^2 \| g_t(\bm y_t) \|^2\!+\!2\alpha_t (\bm \lambda_t-\bm \theta)^\top g_t(\bm y_t), \notag
\end{align}
where we used the non-expansiveness property of the Euclidean projection. Rearranging: 
\begin{align}
\|\bm \lambda_{t+1}\!-\!\bm \theta\|^2\!-\!\|\bm \lambda_t\!-\!\bm \theta\|^2\!\leq\!\alpha_t^2 \| g_t(\bm y_t) \|^2\!+\!2\alpha_t (\bm \lambda_t-\bm \theta)^\top g_t(\bm y_t). \notag
\end{align}
Dividing with $a_t$, setting $\bm \theta\!=\!\bm 0$, and applying the telescopic summation we obtain the final result.

\textbf{Proof of Lemma \ref{lem:viol}:} We have $\bm \lambda_{t+1}=[\lambda_t+a_t g_t(\bm y_t)]^+\succeq \bm \lambda_t+a_t g_t(\bm y_t)$, and dividing by $a_t$ we get:
\[
\frac{\bm \lambda_{t+1}}{a_t} - \frac{\bm \lambda_{t}}{a_t} \succeq g_t(\bm y_t).
\]
Summing telescopically for the first $T$ slots and setting $\bm \lambda_1\!=\!\bm 0$, we obtain:
\[
\sum_{t=1}^T g_t(\bm y_t) \preceq \frac{\bm \lambda_{T+1}}{a_T}+\sum_{t=1}^T \bm \lambda_{t}\left( \frac{1}{a_{t-1}} - \frac{1}{a_t} \right).
\]
Expanding $g_t(\bm y_t)=g(\bm y_t)+\delta_t(\bm y_t)$, dividing with $T$ and taking the norms yields the result.

\textbf{Proof of Lemma \ref{lem:saddle}:} Recall that we defined: $L_t(\bm y, \bm \lambda)=$
\begin{align}
f_t(\bm y)\!+\!\bm \lambda^\top g_t(\bm y)\!=\!f(\bm y)\!+\!\bm \lambda^\top g(\bm y)\!+\!\epsilon_t(\bm y)\!+\!\bm \lambda^\top \delta_t(\bm y).
\end{align} 
Next, we bound the $t$-slot dual function $V_t(\bm \lambda_t) = \min_{\bm y\in \mathcal Y}L_t(\bm y, \bm \lambda_t)$ in terms of the dual function of problem $\mathbb P$, $V(\bm \lambda_t)=\min_{\bm y \in \mathcal Y}f(\bm y) +\bm \lambda_t^\top g(\bm y) $.  Since $\bm y_t \in \arg\min_{\bm y\in \mathcal Y} L_t(\bm y, \bm \lambda_t)$, we have:
\begin{align*}
V_t(\bm \lambda_t) &=  f(\bm y_t) +\lambda_t^\top g(\bm y_t) + \epsilon_t (\bm y_t)+\lambda_t^\top \delta_t(\bm y_t)\\
&\stackrel{(a)}{\le} f(\bm z_t) +\bm \lambda_t^\top g(\bm z_t) + \epsilon_t(\bm z_t)+\bm \lambda_t^\top \delta_t(\bm z_t) \notag \\
&=V(\bm \lambda_t) + \epsilon_t (\bm z_t)+\bm \lambda_t^\top \delta_t(\bm z_t)
\end{align*}
where $(a)$ follows from the minimality of $\bm y_t$.  Hence:
\begin{align*}
f(\bm y^\star)&=V(\bm \lambda^\star)\stackrel{(a)}{\geq} \frac{1}{T} \sum_{t=1}^{T} V(\bm \lambda_t)\\
&\geq\frac{1}{T} \sum_{t=1}^{T} V_t(\bm \lambda_t)-\epsilon_t(\bm z_t) - \bm \lambda_t^\top \delta_t(\bm z_t)\notag \\
&\stackrel{(b)}{\geq} \frac{1}{T} \sum_{t=1}^{T}\Big( L_t(\bm y_t, \bm \lambda_t) - \epsilon_t(\bm z_t) - \bm \lambda_t^\top \delta_t(\bm z_t)\Big)
\end{align*}
where $(a)$ follows from the maximality of $\bm \lambda^\star$ and $(b)$ due to our primal update.

\textbf{Proof of Lemma \ref{lemma:bounded_set}:} $\forall \bm \lambda\!\in\mathcal Q_{v}$ we have: 
\begin{align*}
v\leq V_t(\bm \lambda)\!&=\min_{\bm y\in\mathcal Y} \left\{ f_t(\bm y) +\bm \lambda^\top g_t(\bm y)\right\}\leq f_t(\bm y_s) + \bm \lambda^\top g_t(\bm y_s)\\
&= f_t(\bm y_s) + \sum_{n=1}^{N+1} \lambda_n g_{nt}(\bm y_s)
\end{align*}
Hence it holds: $-\sum_{n=1}^{N+1} \lambda_n g_{nt}(\bm y_s) \le f_t(\bm y_s)-v$. Since $g_{nt}(\bm y_s) < 0$, and $\lambda_n \geq 0$ we get: 
\[\min_n \{-g_{nt}(\bm y_s)\} \sum_{n=1}^{N+1} \lambda_n \leq f_t(\bm y_s) - v \Rightarrow \sum_{n=1}^{N+1} \lambda_n \leq  \frac{f_t(\bm y_s) - v}{q_t}.
\] 
Using that $|f_t(\bm y_s)|\leq \sigma_f$, the definition of $q$ and the fact that $q_t>0$, we arrive at the result.

\textbf{Proof of Lemma \ref{lemma:dual_bound}:} We use an induction argument to show:
\begin{align}\label{eq13}
\|\bm \lambda_t-\bm{\lambda} \|\le \lambda_{max}&:= \frac{2\sigma_f +\|\bm{\lambda}\|\sigma_g}{q} + \frac{ \sigma_g^2}{2q} +\frac{\epsilon}{q}+ \|\bm {\lambda}\|\\ \notag
&+\|\bm {\lambda}_1\|+ a \sigma_g, \,\,\,\forall \bm \lambda \succeq 0.
\end{align}
Trivially, $\|\bm \lambda_1-\bm{\lambda} \|\le\|\bm{\lambda}\|+\|\bm{\lambda}_1\| \le \lambda_{max}$, and assume \eqref{eq13} holds at $t$. We consider two cases.

\textbf{Case} \emph{(i)}: $V_t(\bm \lambda_t) < V_t(\bm{\lambda}) - \frac{a_t \sigma_g^2}{2}$.  Then we can write: 
$
2a_t\big(V_t(\bm \lambda_t) - V_t(\bm{\lambda})\big) < -a_t^2 \sigma_g^2 \,\,\Rightarrow\,\, -2a_t\big(V_t(\bm{\lambda})-V_t(\bm \lambda_t)\big) +a_t^2 \sigma_g^2< 0.
$
Hence, we have:
\begin{align*}
\|\bm \lambda_{t+1}-\bm{\lambda} \|^2 &\leq \|\bm \lambda_t + a_t g_t(\bm y_{t})-\bm{\lambda} \|^2\\
&\leq \|\bm \lambda_t-\bm{\lambda} \|^2 + 2a_t g_t(\bm y_{t})^\top(\bm \lambda_t-\bm{\lambda}) + a_t^2\sigma_g^2 \\
&\stackrel{(a)}{\leq} \|\bm \lambda_t-\bm{\lambda} \|^2 - 2a_t \big(V_t(\bm{\lambda})-V_t(\bm \lambda_t)\big) + a_t^2 \sigma_g^2\\
&\stackrel{(b)}{<} \|\bm \lambda_t-\bm{\lambda} \|^2,
\end{align*}
where $(a)$ follows from the fact that $\bm g_t(y_{t})$ is a subgradient of $V_t(\bm \lambda_t)$ and $(b)$ from the assumptions of the case considered.  Hence, it holds $\|\bm \lambda_{t+1}-\bm{\lambda} \|\le \lambda_{max}$.

\textbf{Case} \emph{(ii)}: $V_t(\bm \lambda_t) \ge V_t(\bm{\lambda}) - \frac{a_t \sigma_g^2}{2}$.
\begin{align*}
\|\bm \lambda_{t+1}\!-\!\bm{\lambda} \|\! &=\! \|[\bm \lambda_t + a_t g_t(\bm y_{t})]^+ -\bm{\lambda}\|
\le \|\bm \lambda_t + a_t g_t(\bm y_{t}) -\bm{\lambda}\|\\
&\leq \|\bm \lambda_t -\bm{\lambda}\| + a_t \sigma_g \\
&\leq \|\bm \lambda_t\| +\|\bm{\lambda}\| + a_t \sigma_g \leq \sum_{n=1}^{N_+1} \lambda_{nt} +\|\bm{\lambda}\| + a_t \sigma_g \\
&\leq \frac{\sigma_f-v}{q}+\|\bm{\lambda}\| + a_t \sigma_g \\
&\leq \frac{\sigma_f}{q} - \frac{1}{q} \Big(\sigma_f+\|\bm \lambda\|\sigma_g -\frac{a_t\sigma_g^2}{2}\Big) +\|\bm{\lambda}\| + a_t \sigma_g   \\
&\leq - \frac{\|\bm \lambda\|\sigma_g}{q} +\frac{a_t\sigma_g^2}{2q} +\|\bm{\lambda}\| + a_t \sigma_g \\
&\leq - \frac{\|\bm \lambda\|\sigma_g}{q} +\frac{a_1\sigma_g^2}{2q} +\|\bm{\lambda}\| + a_1 \sigma_g \!\triangleq\! \lambda_{max} 
\end{align*}
where we used that, by Holders inequality, $\|\bm \lambda_t\|\le \sum_{n=1}^{N+1} \lambda_{nt}$; and applied Lemma \ref{lemma:bounded_set} with:
\[
v=V_t(\bm \lambda) -\frac{a_t\sigma_g^2}{2} \leq \sigma_f+\|\bm \lambda\|\sigma_g -\frac{a_t\sigma_g^2}{2}
\]
and used that $\alpha_1\geq \alpha_t, \forall t$. It follows that $\|\bm \lambda_t-\bm{\lambda} \|\le\lambda_{max}$ and so $\|\bm \lambda_t\|\le\lambda_{max}+\|\bm{\lambda} \|$.
}

\end{document}